\documentclass[conference,compsoc]{commons/IEEEtran}
\usepackage{CJKutf8}
\usepackage{amsmath}
\usepackage{graphicx}
\usepackage{subcaption}
\usepackage{multirow}
\usepackage{fancyvrb}
\usepackage[flushleft, online]{threeparttable}

\ifdefined\ccs
\else
    \usepackage{cite}
\fi
\usepackage{xurl}
\usepackage{array}
\usepackage{algorithmic}
\usepackage{booktabs}
\usepackage{xcolor}
\usepackage{CJKutf8}

\usepackage[normalem]{ulem}
\usepackage{xspace}
\usepackage[disable]{todonotes}
\usepackage{listings}

\usepackage{hyperref}
\usepackage{fancyhdr}

\hypersetup{
    colorlinks=true,
    linkcolor=blue,
    filecolor=magenta,
    urlcolor=cyan,
    pdftitle={Your Title},
    pdfpagemode=FullScreen,
}
\def\revisionDiff{0}

\newcommand{\drevised}[1]{\sout{#1}}

\if\revisionDiff0

    \renewcommand{\drevised}[1]{}
\fi

\newcommand{\hy}[1]{\textcolor{violet}{\textbf{haoyun:} #1}}
\newcommand{\subject}[1]{\noindent\textbf{#1}}
\newcommand{\subsubject}[1]{\noindent\textit{\underline{#1}}}

\newcommand\malurl[1]{\href{notalink}{{\nolinkurl{#1}}}}
\newcounter{finding}
\newcommand{\ignore}[1]{}
\newcommand{\framework}{\texttt{Okara}\xspace} %
\newcommand{\tester}{\texttt{TMV-Hunter}\xspace} %
\newcommand{\analyzer}{\texttt{TMV-ORCA}\xspace} 

\usepackage{tcolorbox}

\newtcolorbox{newpromptbox}[1][]{
    colback=gray!10,
    colframe=gray!50,
    boxsep=5pt,
    arc=4pt,
    fontupper=\ttfamily,
    #1
}

\usepackage{pifont} %

\newcommand{\cmark}{\ding{51}} %
\newcommand{\xmark}{\ding{55}} %

\newcommand{\contribullet}{\noindent\textbf{$\bullet$}\ }

\newcommand{\obs}[1]{\mathbf{o}_{#1}} %
\newcommand{\act}[1]{\mathbf{a}_{#1}} %
\newcommand{\instr}{\mathbf{i}} %
\newcommand{\resp}[1]{\mathbf{r}_{#1}} %
\newcommand{\gui}{\mathbf{G}} %

\begin{document}
\title{Okara: Delving into TLS Certificate Validation Risks in Android Apps with Foundation Models}

\title{Okara: Delving into TLS Certificate Validation Vulnerabilities in Android Apps with Foundation Models}
\title{Okara: Measuring and Diagnosing Pervasive TLS Man-in-the-Middle Vulnerabilities in Android Apps with Foundation Models}
\title{Okara: Detection and Attribution of TLS Man-in-the-Middle Vulnerabilities in Android Apps with Foundation Models}

\newcommand{\AffilOne}{University of Science and Technology of China, Hefei, China}
\newcommand{\AffilTwo}{Ocean University of China, Qingdao, China}
\newcommand{\AffilThree}{Monash University, Melbourne, Australia}

\newcommand{\AuthorOne}{Haoyun Yang}
\newcommand{\AuthorTwo}{Ronghong Huang}
\newcommand{\AuthorThree}{Yong Fang}
\newcommand{\AuthorFour}{Beizeng Zhang}
\newcommand{\AuthorFive}{Junpu Guo}
\newcommand{\AuthorSix}{Zhanyu Wu}
\newcommand{\AuthorSeven}{Xianghang Mi}

\ifdefined\IEEEauthorblockN
    \author{
        \IEEEauthorblockN{
            \AuthorOne\IEEEauthorrefmark{1},
            \AuthorTwo\IEEEauthorrefmark{1},
            \AuthorThree\IEEEauthorrefmark{1},
            \AuthorFour\IEEEauthorrefmark{2},
            \AuthorFive\IEEEauthorrefmark{1},
            \AuthorSix\IEEEauthorrefmark{1} and
            \AuthorSeven\IEEEauthorrefmark{1}\IEEEauthorrefmark{3}
        }
        \IEEEauthorblockA{
            \IEEEauthorrefmark{1}\AffilOne\\
            \IEEEauthorrefmark{2}\AffilTwo\\
            \IEEEauthorrefmark{3}\AffilThree\\
        }
    }
\else
    \author{
        \AuthorOne\inst{1} \and
        \AuthorTwo\inst{1} \and
        \AuthorThree\inst{1} \and
        \AuthorFour\inst{2} \and
        \AuthorFive\inst{1} \and
        \AuthorSix\inst{1} \and
        \AuthorSeven\inst{1,3} \and
    }
    
    \authorrunning{H. Yang et al.}
    
    \institute{
        \AffilOne
        \AffilTwo \and
        \AffilThree
    }
\fi
\makeatother

\fancypagestyle{firstpage}{
  \fancyhf{}
  \renewcommand{\headrulewidth}{0pt}
  \fancyfoot[L]{
    \footnotesize * Corresponding to xianghangmi@gmail.com\\
    * An abridged version of this paper was accepted to ACISP 2026.
  }
  \fancyfoot[C]{\thepage}
}

\IEEEspecialpapernotice {
    \href{https://github.com/ChaseSecurity/Okara}{https://github.com/ChaseSecurity/Okara}
    \vspace{-5mm}
}

\maketitle
\thispagestyle{firstpage}
\pagestyle{plain}

\begin{abstract}
    Transport Layer Security (TLS) is fundamental to secure online communication, yet vulnerabilities in certificate validation that enable Man-in-the-Middle (MitM) attacks remain a pervasive threat in Android apps. Existing detection tools are hampered by low-coverage UI interaction, costly instrumentation, and a lack of scalable root-cause analysis. We present Okara, a framework that leverages foundation models to automate the detection and deep attribution of TLS MitM Vulnerabilities (TMVs). Okara's detection component, TMV-Hunter, employs foundation model-driven GUI agents to achieve high-coverage app interaction, enabling efficient vulnerability discovery at scale. Deploying TMV-Hunter on 37,349 apps from Google Play and a third-party store revealed 8,374 (22.42\%) vulnerable apps. Our measurement shows these vulnerabilities are widespread across all popularity levels, affect critical functionalities like authentication and code delivery, and are highly persistent with a median vulnerable lifespan of over 1,300 days. Okara's attribution component, TMV-ORCA, combines dynamic instrumentation with a novel LLM-based classifier to locate and categorize vulnerable code according to a comprehensive new taxonomy. This analysis attributes 41\% of vulnerabilities to third-party libraries and identifies recurring insecure patterns, such as empty trust managers and flawed hostname verification. We have initiated a large-scale responsible disclosure effort and will release our tools and datasets to support further research and mitigation.

\end{abstract}
\section{Introduction}
\label{sec:intro}

Transport Layer Security (TLS) ensures confidentiality and integrity of internet communication~\cite{holz2020tracking, lee2021tls, mankowski2023tls}. However, its security relies on correct client-side certificate validation, a complex process involving chain verification and hostname checks that remains notoriously error-prone~\cite{greenwood2014smv, pourali2024racing}. These implementation errors lead to TLS MitM Vulnerabilities (TMVs), allowing attackers to intercept encrypted traffic. Unlike modern web browsers, the Android ecosystem remains highly susceptible. As detailed in Section~\ref{sec:tls_analysis}, TMVs represent a pervasive threat, affecting apps across all popularity levels and jeopardizing sensitive user data.

Addressing this threat requires tools capable of large-scale detection and in-depth analysis. However, the current landscape of TMV detection tools is inadequate for a modern, large-scale study. Existing tools~\cite{10.1145/2382196.2382205, greenwood2014smv, onwuzurike2015danger, wang2019dcdroid, pourali2024racing} face several critical limitations: (1) \textit{Ineffectiveness and Obsolescence}: Many proposed tools are either closed-source or no longer maintained, rendering them incompatible with contemporary Android versions and app datasets. (2) \textit{Limited Detection Coverage}: Static analysis tools are typically error-prone and are subject to a low coverage of critical TLS flows. On the other hand, dynamic analysis tools rely on UI interaction to trigger TLS traffic, but their interaction components often lack the intelligence to achieve high coverage, failing to explore deep app functionalities and thus missing vulnerable TLS flows. (3) \textit{Lack of Scalable Origin and Root Cause Analysis (ORCA)}: Once vulnerabilities are detected, previous works either forgo root cause analysis or rely on manual, labor-intensive methods that do not scale to thousands of apps. The taxonomies used for categorizing errors are often coarse-grained, providing limited insight for effective remediation. (4) \textit{Poor Reproducibility and Efficiency}: Methodological constraints, such as dependencies on specific device models or root privileges, coupled with inefficient analysis pipelines (e.g., taking several minutes per app), hinder reproducibility and large-scale deployment.

To overcome these challenges, we introduce \framework, a comprehensive framework that leverages recent advances in foundation models to automate the detection and root-cause analysis of TMVs in Android apps. \framework is designed as a toolchain with two core components, each addressing a distinct research goal.

First, to enable large-scale and up-to-date detection of TMVs, we develop \tester, a dynamic analysis tool for efficient TMV detection. A key innovation of \tester is its use of foundation model-driven GUI agents. These agents understand natural language instructions and app context, enabling them to perform realistic, high-coverage interactions with app interfaces. This approach far surpasses the random or rule-based strategies of prior tools, triggering a more diverse range of TLS traffic for testing. \tester is designed to be parametrizable and resource-efficient, making it feasible to scan tens of thousands of apps.

Second, to understand the origins and root causes of the detected vulnerabilities, we develop \analyzer. This tool automates the traditionally manual process of root cause analysis. It combines dynamic instrumentation to precisely locate the vulnerable code snippets within apps with a novel Large Language Model (LLM)-based classifier. This classifier categorizes the vulnerable code according to a new, comprehensive, and fine-grained taxonomy of TLS validation flaws that we developed. \analyzer systematically attributes vulnerabilities to either app developers or third-party libraries, providing crucial data for targeted disclosure and quantifying the significant role of third-party code in propagating these security flaws.

We deploy \tester on a large dataset of 37,349 apps from Google Play and a major third-party app store (AppChina), identifying 8,374 (22.42\%) vulnerable apps. Our measurement study (Section \ref{sec:tls_analysis}) reveals that these vulnerabilities are widespread across all app categories and popularity levels, affect critical functionalities like authentication and code delivery, and are highly persistent—with a median vulnerable lifespan of over 1,300 days. Using \analyzer, we locate and categorize vulnerable code from thousands of apps, finding that 41\% of these vulnerabilities originate from third-party libraries, and identifying recurring insecure patterns such as empty trust managers and flawed hostname verification.

Based on our findings, we have initiated a large-scale responsible disclosure effort to notify the developers and library maintainers of the affected apps. We are committed to releasing \framework, including \tester and \analyzer, along with our datasets, to the research community to facilitate further research and mitigation efforts.

\subject{Contribution.} In summary, this paper makes the following contributions:

\contribullet We present \framework, a framework for the large-scale detection and in-depth root cause analysis of TLS MitM vulnerabilities in Android apps.

\contribullet We introduce \tester, a novel dynamic analysis tool that leverages foundation model-based GUI agents to achieve high-coverage, realistic user interaction, enabling efficient and scalable vulnerability discovery that surpasses prior methods.

\contribullet We conduct the most extensive measurement study of TMVs to date on 37,349 Android apps, providing a multi-faceted view of their prevalence, criticality, and longitudinal evolution.

\contribullet We develop \analyzer, an automated root cause analysis tool that combines dynamic instrumentation with an LLM-based classifier to locate vulnerable code and categorize errors according to a new, comprehensive taxonomy.

\contribullet We perform a systematic attribution of vulnerabilities to app developers or third-party libraries, quantifying the significant role of the latter, and initiate a large-scale responsible disclosure campaign.

\subject{Ethical Considerations.}
Our research adhered to strict ethical standards. All testing, including Man-in-the-Middle (MitM) attacks, was confined to a controlled laboratory environment on our own devices. We did not intercept or compromise any real user data. The analyzed apps were obtained from public sources, while resulting sensitive vulnerability information is kept confidential. We have initiated a responsible disclosure process to notify affected developers and library providers, giving them an opportunity to address the vulnerabilities before any potential public release of identifying details, as detailed in Section~\ref{sec:disclosure}.

\subject{Paper Organization.} The remainder of this paper is organized as follows. Section~\ref{sec:pre} discusses background and related work. Section~\ref{sec:tls_detection} details the design and evaluation of \tester. Section~\ref{sec:tls_analysis} presents our large-scale measurement study of detected TMVs. Section~\ref{sec:tls_causality} describes \analyzer and our root cause analysis. Section~\ref{sec:disclosure} outlines our responsible disclosure effort, and Section~\ref{sec:discussion} discusses limitations and concludes.

\section{Preliminaries}
\label{sec:pre}
\subject{TLS Certificate Validation Risks in Android Apps.} Since the inception of the Android ecosystem, a substantial body of research has focused on addressing the security challenges associated with TLS certificate validation failures. Despite these efforts and the continuous security enhancements within the Android ecosystem, this study reveals that this long-existing issue remains prevalent to a significant extent.

Among these works, Sascha et al.~\cite{10.1145/2382196.2382205} pioneered the use of static analysis with their tool, MalloDroid, which identifies apps potentially vulnerable to man-in-the-middle attacks. However, static analysis turns out to be error-prone\cite{greenwood2014smv}. To improve accuracy while managing computational costs, Sounthiraraj et al.~\cite{greenwood2014smv} introduced Smv-hunter, which combines static analysis to filter potentially vulnerable apps with dynamic execution to confirm actual vulnerabilities. 

Nevertheless, Smv-hunter and subsequent dynamic analysis works~\cite{onwuzurike2015danger, wang2019dcdroid, pourali2024racing}  share a common set of limitations. Specifically, their UI interaction components lack the intelligence to effectively mimic real-world user interactions, leading to limited coverage of TLS traffic flows. Furthermore, for Android apps with TLS vulnerabilities, these studies either fail to conduct root cause analysis or rely on manual, labor-intensive methods that are not scalable for analyzing a large volume of Android apps. Besides, previous studies also fall short in a clear and fine-grained root cause taxonomy.  

Then, Pourali et al.~\cite{pourali2024racing} made further efforts to achieve fine-grained attribution, i.e., attributing a validation failure to the responsible party, which could be either the app developer or a third-party library developer. However, their methodology requires root privileges and extensive instrumentation, making it inefficient and impractical for analyzing a large volume of Android apps. Moreover, such heavy instrumentation can trigger abnormal behaviors in apps that detect instrumentation, such as crashing or failing to generate network traffic, thereby reducing the coverage of TLS validation failures.

Another significant limitation lies in the reproducibility of prior works, which hinders direct empirical comparisons. Many of the tools developed in these studies are either closed-source or no longer maintained, making them incompatible with modern Android versions and contemporary app datasets. Additionally, the lack of publicly available evaluation datasets prevents replication or comparative analysis on a shared benchmark. Methodological constraints, such as dependencies on specific physical device models and the need for root privileges, further restrict the scalability of reproducing these experiments. These challenges highlight the need for a reproducible and scalable baseline, which we address in this study through {\framework}.

To address these limitations, this study introduces several novel techniques. Particularly, We utilize multimodal foundation models and advanced open-source GUI agents to automate interactions with Android apps, achieving broader coverage of TLS traffic flows (Section~\ref{sec:tls_detection}). Besides, our approach separates the detection of TLS validation failures from root cause analysis, enabling efficient high-throughput detection (\tester) and detailed in-depth analysis (\analyzer). Additionally, we propose an automated method to classify TLS validation failures into a hierarchical and fine-grained taxonomy of implementation errors, ensuring precise attribution. Experimental results validate the effectiveness of our approach, demonstrating high performance in both detection and root cause analysis (Sections~\ref{sec:tls_detection} and~\ref{sec:tls_causality}).

As our {\tester} relies on GUI agents to automate Android GUI traversal, we present relevant concepts and background knowledge as below.

\subject{GUI Agents.} GUI agents are autonomous systems designed to process language instructions and interact with software systems through their graphical user interfaces (GUIs). Formally, a GUI agent can be defined as: 
\[
    \gui \big(\instr, \obs{1}, \act{1}, \obs{2}, \act{2}, \dots, \obs{i-1}, \act{i-1}, \obs{i}\big) 
    \rightarrow \big( \act{i}, \resp{i} \big)  
\]
Here, $\gui$ represents the GUI agent, $\instr$ denotes the language instruction specifying a GUI task (e.g., booking a flight), $\obs{i}$ represents the observation at step $i$, $\act{i}$ denotes the action taken at step $i$, and $\resp{i}$ represents the textual response at step $i$ in addition to the action tokens, such as reasoning or "thinking" tokens~\cite{qin2025uitarspioneeringautomatedgui}. 

The nature of observations ($\obs{i}$) varies across GUI agents. Observations can consist of pure visual elements, such as screenshots of the current screen or software interface, or they may include structural information, such as the layout of the GUI (e.g., an HTML layout). Similarly, the action space ($\act{i}$) encompasses standard GUI operations, such as clicking buttons or updating input fields with predefined information (e.g., entering a customer's email address or specifying the quantity of goods).

Some GUI agents~\cite{zheng2024gpt4visiongeneralistwebagent} are designed as modular systems that integrate a series of expert models through prompts and bridging code. In contrast, other agents are implemented as end-to-end multimodal foundation models, utilizing a single model to handle tasks such as perception, reasoning, memory, and action execution. These are referred to as native GUI agents~\cite{qin2025uitarspioneeringautomatedgui,lin2024showui, xu2024aguvisunifiedpurevision, wu2024osatlasfoundationactionmodel}. Compared to their modular counterparts, native GUI agents demonstrate superior generalizability and adaptability while significantly reducing manual efforts, such as rigid workflow design and labor-intensive prompt engineering. Consequently, this study adopts native GUI agents to enable comprehensive and automated interaction with Android apps, facilitating the dynamic detection of TLS validation failures with enhanced efficiency and accuracy (Section~\ref{subsec:tester_design}).

Next, we move to present \tester, a core module of our \framework framework, designed to realistically interact with Android apps and dynamically detect TLS MitM vulnerabilities. 

\section{Detecting TLS MitM Vulnerabilities}
\label{sec:tls_detection}
To address our first research question—gaining an up-to-date understanding of TLS validation risks in Android apps—we require a detection tool capable of efficiently scanning large numbers of apps and accurately identifying validation failures.
However, existing tools suffer from several limitations, including inefficiency, incompatibility with recent Android versions, and lack of openness, etc. Table~\ref{tab:tls_detection_comparison} summarizes four dynamic analysis tools for detecting TLS validation failures in Android apps. The first three tools are either closed source or no longer maintained, rendering them incompatible with current Android systems and applications. Marvin, while open source and actively maintained as of 2024, requires extensive app instrumentation to correlate vulnerable TLS traffic with the responsible app code. This process is time-consuming, with the original study~\cite{pourali2024racing} reporting an average analysis time of 3 to 12 minutes per app. Similarly, the other three tools also exhibit unsatisfying efficiency, limiting their detection scale to only thousands of apps.

These limitations highlight the need for a more efficient, scalable, and up-to-date solution for detectingx TLS MitM vulnerabilities in modern Android applications. Therefore, we develop the \textbf{T}LS \textbf{M}itM \textbf{V}ulnerability Hunter—\tester for short—to address these gaps. Next, we highlight the design, implementation,  evaluation, and deployment of \tester.
\begin{table}
\centering
\caption{Comparison of Dynamic Analysis Tools for Detecting TLS Validation Failures in Android Applications.}
\label{tab:tls_detection_comparison}
\begin{threeparttable}
\begin{tabular}{lcccc}
\toprule
\textbf{Tool} & \textbf{Open} & \textbf{Compatibility\tnote{a}} & \textbf{Efficiency} & \textbf{Root\tnote{b}} \\
\midrule
SMV-Hunter~\cite{greenwood2014smv} & \cmark & No & Low & No \\
SSLDetector~\cite{tang2019ssldetecter} & \xmark & No & Low & No\\
DCDroid~\cite{dcdroid} & \xmark & No & Medium & No\\
Marvin~\cite{pourali2024racing} & \cmark & Yes & Low & Yes\\
\bottomrule
\end{tabular}
\begin{tablenotes}
\footnotesize
\item[a] Indicates whether the tool is actively maintained and compatible with recent Android versions.
\item[b] Specifies if the tool requires root privileges on the device.
\end{tablenotes}
\end{threeparttable}
\end{table}
\begin{figure*}
    \centering
    \includegraphics[width=\textwidth]{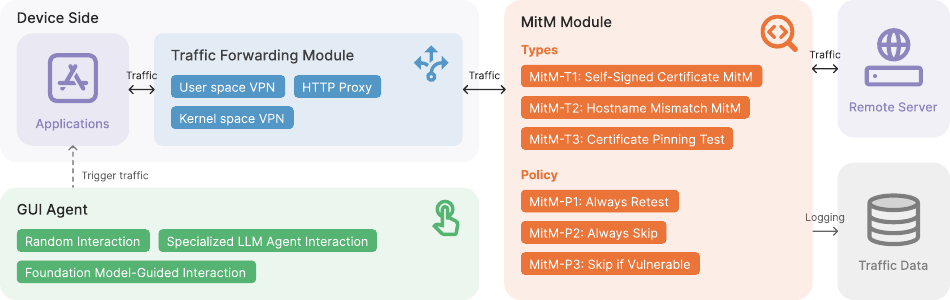}
    \caption{The design of \tester.}
    \label{fig:tester_design}
\end{figure*}
\subsection{Design of {\tester}}
\label{subsec:tester_design}
The primary objective of \tester is to automatically and efficiently detect TLS certificate validation failures in Android applications via dynamic analysis. As illustrated in Figure~\ref{fig:tester_design}, \tester accepts as input an Android application and a set of configurable testing parameters (such as time budget and the maximum number of GUI interaction steps). It then systematically exercises the app, employing realistic and high-coverage GUI interactions to stimulate a diverse range of TLS traffic. Simultaneously, \tester performs practical man-in-the-middle (MitM) attacks to intercept, test, and analyze all TLS flows generated during execution. The output is a comprehensive report detailing all detected TLS validation failures, including the associated traffic flows and relevant metadata. Notably, \tester is designed to be fully automated and standalone, enabling scalable and accurate detection of TLS validation vulnerabilities in modern Android apps.

\tester is architected around three core modules: a GUI agent, a traffic forwarding module, and a MitM module. Each module incorporates one or more enhancements over prior solutions. The GUI agent supports multiple interaction strategies—including advanced LLM-based approaches—to achieve high-coverage and realistic user interactions. The traffic forwarding module ensures that only the target app's traffic is securely routed for MitM testing, without impacting the system or other co-located Android apps. The MitM module simulates realistic attack scenarios, focusing on practical MitM attacks against TLS flows. Below, we detail the design and roles of these modules.

\subject{The GUI Agent.} Detecting TLS validation failures requires triggering representative TLS traffic flows, which in turn depends on realistic and thorough interaction with app GUIs. Our GUI agent is designed to simulate real user behavior by systematically traversing UI elements and performing common actions such as click, type, and scroll.

The main challenge is selecting which action to perform on which UI element to maximize TLS flow coverage. Previous tools relied on random or rule-based strategies, which lack context awareness and realism. To overcome this, our agent leverages foundation models—including vision-language and large language models (VLMs/LLMs)—to understand natural language interaction instructions, interpret app context and guide interactions, enabling high-coverage and realistic user simulation.

The agent supports three interaction strategies: (1) \textit{random}, which selects actions and elements randomly; (2) \textit{general LLM}, which uses general-purpose foundation models for context-aware actions; and (3) \textit{specialized LLM}, which employs models fine-tuned for GUI tasks to further improve coverage and realism. Implementation details are detailed in \S\ref{subsec:tester_impl}, and a comparative evaluation is given in \S\ref{subsec:tester_eval}.

\subject{The Traffic Forwarding Module.} This module is responsible for routing all network traffic from the target Android app through the MitM module for testing. The design goals are: (1) to avoid requiring root privileges, enabling deployment of \tester on real, unrooted devices; (2) to selectively forward only the traffic from designated apps, ensuring other apps on the device are unaffected; and (3) to support forwarding of both UDP and TCP traffic, thereby covering UDP-based TLS flows such as HTTP/3 in addition to traditional HTTPS and WebSocket traffic. Achieving these goals allows \tester to be deployed flexibly and ensures comprehensive coverage of diverse TLS traffic flows.

To fulfill these requirements, the module leverages Android's \texttt{VpnService} to establish a VPN service, which routes all network traffic originating from the target app to the MitM node (i.e., the VPN server), as well as relaying responses back to the app. By utilizing Android's per-app VPN APIs~\footnote{\url{https://developer.android.com/develop/connectivity/vpn\#allowed-apps}}, the traffic forwarding module can configure an explicit allowlist, ensuring that only traffic from designated apps is routed through the VPN, while all other apps and system processes continue to use the device's default networking stack. This selective forwarding approach provides fine-grained control over the analysis scope and preserves normal device functionality for non-targeted apps.

\subject{The MitM Module.} The MitM module is responsible for  conducting practical man-in-the-middle (MitM) tests on TLS traffic flows. This module consists of two main components: a VPN server that receives traffic from the traffic forwarding module and relays responses back, and a MitM test suite that detects real-world TLS certificate validation vulnerabilities.

The MitM test suite is designed to systematically identify three primary categories of TLS validation vulnerabilities. The first category, denoted as \textit{MitM-T1}, simulates an attacker leveraging an untrusted root certificate authority (CA). In this scenario, during the TLS handshake, the attacker presents a leaf certificate that is otherwise valid but is signed by a self-generated root CA that is not trusted by the Android device. This test reveals whether the app properly enforces the trust anchor validation required by the platform.

The second category, \textit{MitM-T2}, focuses on domain name mismatches. Here, the attacker presents a certificate issued by a widely trusted CA, but for a domain name under the attacker's control rather than the intended server. This scenario tests whether the app correctly verifies that the certificate's subject matches the expected domain, thereby preventing impersonation attacks even when the certificate chain is otherwise valid.

The third category, \textit{MitM-T3}, is designed to assess the enforcement of certificate pinning. In this case, the attacker installs their own root certificate into the device's CA certificate store, making it trusted by the system. The MitM attack will succeed if the target app fails to implement certificate pinning and thus accepts certificates signed by any trusted CA, including those added by an attacker. This test is crucial for evaluating whether apps are resilient against attacks that exploit compromised or maliciously added root certificates.

To ensure comprehensive testing, the MitM module tracks previously tested traffic destinations, identified by their fully qualified domain names. When a new TLS flow to a previously encountered destination is detected, the module applies one of three configurable retesting policies: (1) \textit{MitM-P1}, where the MitM test is performed on all TLS flows, regardless of whether the destination has been previously tested; (2) \textit{MitM-P2}, where the MitM test is conducted only once for each unique FQDN, and subsequent flows to the same destination are skipped; and (3) \textit{MitM-P3}, where tests are repeated for a given FQDN until a vulnerability is detected, after which subsequent flows to that destination are skipped. This flexible design allows the balance between thoroughness and efficiency according to specific testing needs.

\subject{Key Parameters of \tester.} Given the design of core modules introduced, 
we now formally define the key parameters that govern the operation of {\tester}:

\begin{itemize}
    \item \textit{GUI Interaction Strategy} ($S_{\mathrm{GUI}}$): Specifies the approach used for interacting with the app. Options include (1) random interaction, (2) interaction guided by general-purpose foundation models, and (3) interaction using a specialized LLM-based GUI agent.
    \item \textit{Interaction Step Budget} ($N_{\mathrm{steps}}$): The maximum number of GUI actions (e.g., clicks) allowed during testing. $N_{\mathrm{steps}} = -1$ indicates no limit.
    \item \textit{Time Budget} ($T_{\mathrm{max}}$): The maximum allowed duration (in seconds) for each testing session.
    \item \textit{Interaction Wait Time} ($T_{\mathrm{wait}}$): The time interval (in seconds) to wait between two consecutive GUI interaction steps.
    \item \textit{MitM Retest Policy} ($P_{\mathrm{MitM}}$): Determines how \tester handles new TLS flows to destinations already tested. Options include: always retest, always skip, or skip only if a previous test for that destination revealed a MitM vulnerability.
\end{itemize}

\subsection{Implementation of \tester}
\label{subsec:tester_impl}

Building on the design of \tester, we now describe the key implementation details.

\subject{Implementation of the GUI Agent.} To enable automated interaction with UI elements in target apps, we initially considered Appium, a widely used open-source UI automation framework. Appium facilitates programmable control of both web browsers and mobile devices via the WebDriver protocol. For Android automation, Appium relies on the platform's native accessibility APIs through integration with the UiAutomator2 library, providing robust and flexible UI manipulation capabilities.

However, our evaluation identified two significant limitations with Appium. First, it introduces considerable latency due to its multi-layered communication architecture—spanning the test script, WebDriver client, UiAutomator2 client, and UiAutomator2 server—which slows down UI interactions. Second, Appium is prone to frequent exceptions and errors, which can disrupt automated workflows and hinder large-scale deployment.

To address these issues, we implemented the GUI agent directly using the UiAutomator2 library~\footnote{\url{https://github.com/openatx/uiautomator2}}. This approach eliminates unnecessary communication layers, resulting in faster and more reliable UI interactions with significantly fewer exceptions. By leveraging UiAutomator2 natively, our GUI agent achieves both higher efficiency and greater robustness, making it well-suited for scalable, automated testing in \tester.

For generating the next UI action, our GUI agent supports three interaction strategies, all implemented as a unified Python test script. Two of these strategies require efficient deployment of foundation models. To this end, we utilize the vLLM framework to deploy the selected foundation model as a dedicated inference server, which can handle concurrent inference requests via the API Endpoint.

For the two LLM-based strategies, we employ \textit{Qwen2.5-VL-32B-Instruct}~\cite{Qwen2.5-VL} as the general-purpose foundation model and \textit{UI-TARS-1.5-7B}~\cite{qin2025ui} as the specialized model. Notably, the general-purpose model has 32 billion parameters, while the specialized model has only 7 billion. This choice is informed by preliminary experiments, which revealed that smaller variants of Qwen2.5-VL struggled to reliably compute correct coordinates from UI element bounds and consistently adhere to the specified action output format.

Although both models are provided with the same task instructions, their prompt structures differ. The general-purpose LLM is prompted with a single, comprehensive user message that consolidates all contextual information, including the task description, action space, format requirements, interaction history, and current screen state. In contrast, the specialized LLM (UI-TARS) employs a multi-turn conversational format aligned with its training methodology. This format includes a system message, the interaction history, and the current GUI screenshot as the final user message to prompt the next action. Detailed prompt templates for both approaches are provided in Appendix~\ref{appendix:prompts}.

\subject{Implementation of Traffic Forwarding and MitM.}
The traffic forwarding module is implemented using a WireGuard VPN client on the Android device and a WireGuard VPN server co-located with the MitM module. For the client, we select Android GUI for WireGuard, an open-source solution that fully supports Android's per-app VPN APIs. This enables \tester to selectively forward traffic from only the target apps, ensuring that network activity from other apps and the system remains unaffected.

Currently, the VPN client is implemented to support two modes. One is a userspace VPN service, which does not require root privileges and is thus suitable for unrooted devices, though it may be less stable (e.g., can be killed by the system) and may be detected by some applications. It supports selective traffic forwarding. The other mode runs the VPN service in the kernel space, which requires root privileges and will route traffic of all apps and the sytem, but offers greater stability and less interference with app behavior.

On the server side, we utilize the WireGuard VPN server integrated with the mitmproxy framework. This design allows seamless co-deployment of the VPN server and the MitM module, simplifying the overall architecture and reducing operational overhead.

The MitM test suite itself is implemented as a custom add-on for mitmproxy. In addition to conducting the three types of MitM tests (Section~\ref{subsec:tester_design}), this add-on enables \tester to record detailed metadata for each TLS traffic flow, along with the corresponding MitM test results. This comprehensive logging facilitates thorough analysis and reproducibility of test results.

\subsection{Evaluation of \tester}
\label{subsec:tester_eval}

With the implementation of \tester complete, we conduct a comprehensive evaluation to assess its performance and to identify optimal deployment strategies that balance efficiency and effectiveness. Our evaluation proceeds in two main stages. First, we systematically compare different configurations of the GUI agent module, profiling their capabilities in UI element traversal and TLS traffic triggering. Second, we assess the end-to-end performance of \tester in detecting MitM vulnerabilities. These evaluation tasks underscore the necessity of an annotated dataset comprising representative Android apps, essential UI elements and TLS traffic flows for each app, as well as vulnerable TLS flows confirmed to be susceptible to MitM vulnerabilities. Next, we move to present this evaluation dataset. 

\subject{The Evaluation Dataset.} To construct this dataset, we select the top \textit{100} most popular Android apps from the third-party app store AppChina, using download volume as a proxy for popularity. For each app, a researcher performs extensive manual interaction for 50 interaction steps as counted by the accessibility framework. This interaction process is instructed to focus on essential UI elements and attempting to trigger critical traffic flows. This process yields, for each app, a \textit{ground truth} set of distinct Android activities and TLS traffic flows, which serve as the basis for evaluating the GUI agent. 
In total, the resulting manual interaction dataset comprises 567 unique UI elements (Android activities) and 2,341 distinct TLS traffic flows. 
For UI elements, we limit the granularity to Android Activity (i.e., an window) and consider two elements to be duplicates if they share both the same Android app identifier and the same activity name. Similarly, two TLS traffic flows are treated as duplicates if they originate from the same app and target the same TLS server address (i.e., the same fully qualified domain name). 

Additionally, we perform comprehensive manual MitM testing to assess whether each app and its corresponding TLS flows are susceptible to the defined MitM attacks. These confirmed vulnerabilities serve as the ground truth for evaluating the end-to-end detection performance of \tester. The resulting manual MitM dataset consists of 69 vulnerable apps (encompassing 487 vulnerable TLS flows and 2,372 non-vulnerable TLS flows) and 31 non-vulnerable apps (with 532 non-vulnerable TLS flows). The deduplication strategy described above is consistently applied when enumerating TLS traffic flows to ensure accurate and non-redundant statistics.

\subject{Computing Setup.} All evaluation experiments are conducted on cloud servers equipped with ARM architecture, primarily utilizing Alibaba Cloud ECS (g6r.2xlarge, Ampere Altra) and AWS EC2 (c6g.2xlarge, AWS Graviton2). The Android environment is provisioned using redroid~\footnote{\url{https://github.com/remote-android/redroid-doc}}, which enables scalable deployment of recent Android versions in the cloud, ensuring compatibility with the latest app releases. Foundation model inference services are hosted separately on dedicated  servers equipped with three high-performance GPUs, supporting efficient and concurrent model inference. This infrastructure enables robust, large-scale automated testing and rapid model inference.

\newcommand{\UIcoverage}{C_{\mathrm{UI}}}
\newcommand{\UInewcoverage}{C^{novel}_{\mathrm{UI}}}
\newcommand{\TLScov}{C_{\mathrm{TLS}}}
\newcommand{\TLSnewcov}{C^{novel}_{\mathrm{TLS}}}
\newcommand{\UIauto}{\mathcal{E}_{\mathrm{auto}}}
\newcommand{\UImanual}{\mathcal{E}_{\mathrm{manual}}}
\newcommand{\TLSauto}{\mathcal{S}_{\mathrm{auto}}}
\newcommand{\TLSmanual}{\mathcal{S}_{\mathrm{manual}}}

\begin{table}
\centering
\footnotesize
\caption{Comparison of Different GUI Agent Setups}
\label{tab:gui_agent_coverage}
\begin{threeparttable}
\begin{tabular}{lcccccc}
\toprule
\textbf{Strategy} & WT\tnote{a} & $\UIcoverage$ & $\UInewcoverage$ & $\TLScov$ & $\TLSnewcov$ & TC\tnote{b} \\
\midrule
Random           & 2 & 0.4056 & \textbf{0.3704} & 0.4689 & 0.2691 & \textbf{95.20} \\  
Random           & 4 & 0.4021 & 0.3545 & 0.4638 & 0.2666 & 158.69 \\  
Random           & 8 & 0.3845 & 0.3263 & 0.4805 & 0.3039 & 271.57 \\  
\midrule  
G-LLM\tnote{c}  & 4 & 0.3704 & 0.1499 & 0.5266 & 0.3069 & 531.76 \\  
G-LLM            & 8 & 0.3651 & 0.1129 & 0.5387 & 0.2691 & 679.05 \\  
G-LLM            & 16 & 0.3492 & 0.1164 & 0.5160 & 0.2610 & 1051.46 \\  
\midrule  
S-LLM\tnote{c}  & 4 & 0.3651 & 0.1852 & 0.5405 & 0.2762 & 333.55 \\  
S-LLM            & 8 & 0.3386 & 0.1464 & 0.5112 & 0.2456 & 397.66 \\  
S-LLM            & 16 & 0.3580 & 0.1693 & 0.5468 & 0.2979 & 689.39 \\  
\midrule
SOTA~\tnote{d}   & 4 & \textbf{0.4515} & 0.2945 & \textbf{0.5894} & \textbf{0.4449} & 1014.97 \\
\bottomrule
\end{tabular}
\begin{tablenotes}
\item [a] WT denotes the interaction wait time between consecutive actions, measured in seconds.
\item [b] TC denotes the average time cost for each strategy, measured in seconds.
\item [c] G-LLM refers to the use of a general-purpose large language model for GUI interaction, while S-LLM denotes a specialized LLM fine-tuned for GUI-related tasks.
\item [d] SOTA denotes a state-of-the-art (SOTA) general LLM adopted in our evaluation, namely, \textit{Gemini 2.5 Flash}.
\end{tablenotes}
\end{threeparttable}
\end{table}

\subject{Evaluation of the GUI Agent.} To evaluate effectiveness of the GUI Agent, we adopt two metrics to profile UI element traversal and another two for profiling the triggering of TLS traffic, which are elaborated as below: 

First, \textit{UI Element Coverage} (\(\UIcoverage\)) measures the ratio of unique UI elements (Android activities) triggered by the GUI agent that were also observed during manual testing, defined as \(\UIcoverage = \frac{|\UIauto \, \cap \, \UImanual|}{|\UImanual|}\). Second, the \textit{New UI Element Ratio} (\(\UInewcoverage\)) quantifies the proportion of UI elements triggered by the GUI agent that were not observed during manual testing, given by \(\UInewcoverage = \frac{|\UIauto \, \setminus \, \UImanual|}{|\UIauto|}\). Third, \textit{TLS Server Coverage} (\(\TLScov\)) captures the proportion of unique TLS servers (FQDNs) triggered by the GUI agent that were also observed during manual testing, calculated as \(\TLScov = \frac{|\TLSauto \, \cap \, \TLSmanual|}{|\TLSmanual|}\). Finally, the \textit{New TLS Server Ratio} (\(\TLSnewcov\)) represents the fraction of TLS servers triggered by the GUI agent that were not observed in manual testing, defined as \(\TLSnewcov = \frac{|\TLSauto \, \setminus \, \TLSmanual|}{|\TLSauto|}\).

Given these metrics, we conduct a series of experiments to evaluate the effectiveness of different GUI agent strategies and interaction wait times. In these experiments, we set up the maximum interaction steps to 50, thus being consistent with the manual testing procedure. For each strategy, we vary the interaction wait time between consecutive actions to observe its impact on coverage metrics. For the LLM-based interaction strategies, we employ Qwen2.5-VL-32B-Instruct as the general-purpose foundation model and UI-TARS-1.5-7B as the specialized LLM for GUI tasks. 
Also, the action space of all strategies are consistently set up to include the following 8 actions: 

\begin{itemize}
    \item \texttt{click [x] [y]}: Click a location on the screen.
    \item \texttt{long\_click [x] [y]}: Long-click a location on the screen.
    \item \texttt{type [text]}: Type the text in the currently focused input box.
    \item \texttt{scroll [x] [y] [direction]}: Slide in the specified position and direction, direction can be `up', `down', `left', `right'.
    \item \texttt{drag [start\_x] [start\_y] [end\_x] [end\_y]}: Drag from (start\_x, start\_y) to (end\_x, end\_y).
    \item \texttt{back}: Press the return key.
    \item \texttt{wait}: Wait for a period of time, often used to wait for a page to load.
    \item \texttt{finish}: End the test early, only if the task can't be completed anyway.
\end{itemize}

\subsubject{Efficiency of the GUI Agent.} To profile efficiency of the GUI agent, we report the average time cost for each interaction strategy, measured in seconds. This metric reflects the total duration required by the GUI agent to complete an interaction session, encompassing both the execution time of each action and the configured wait intervals between actions. Evaluating this metric is essential for understanding the practical trade-offs between coverage and efficiency across different interaction strategies.

As shown in Table~\ref{tab:gui_agent_coverage}, the Random strategy is the most time-efficient, completing sessions in the shortest duration. In contrast, the General LLM strategy incurs the highest time cost, primarily due to the substantial computational overhead associated with LLM inference. Notably, the Specialized LLM strategy achieves a favorable balance, offering significantly lower time cost than the General LLM while maintaining competitive coverage. This efficiency gain can be attributed to the smaller model size of the Specialized LLM (7 billion parameters) compared to the General LLM (32 billion parameters), resulting in faster inference and reduced computational demands.

\subsubject{Impact of LLM-Related Parameters.} We further assess how LLM-specific parameters affect GUI agent performance. We focus on three aspects: (1) \textit{LLM model type}—comparing general-purpose versus specialized models; (2) \textit{LLM input modality}—whether the model receives only structured UI element data or also incorporates GUI window screenshots; and (3) \textit{action space}—whether the LLM selects from all possible actions or is restricted to a subset (e.g., only click and swipe).

Detailed results are provided in Table~\ref{tab:llm_setups_wide} of Appendix~\ref{appendix:llm_params_eval}. In summary, our evaluation yields three key findings: (1) The underlying LLM's capability is a decisive factor—state-of-the-art general-purpose models such as Gemini 2.5 Flash consistently outperform other LLMs in both UI element coverage and TLS traffic coverage. (2) Incorporating visual context through GUI window screenshots further enhances UI coverage, although it may slightly reduce the discovery of new TLS traffic flows. (3) Restricting the action space to only \textit{click} and \textit{back} actions results in marginally lower, but still comparable, performance for both UI exploration and TLS traffic triggering. These insights inform our choice of LLM configuration for subsequent large-scale experiments.

Based on these findings, unless specialized otherwise, we configure the GUI agent in subsequent experiments to use the specialized LLM (namely, UI-TARS-1.5-7B) with screenshot input and full action space. The interaction wait time is set to 4 seconds to balance coverage and efficiency.

\subject{End-to-End Performance of \tester.} Four metrics have been employed to measure the end-to-end performance of \tester with regards to detection of MitM vulnerabilities. 

\newcommand{\Appdet}{\mathcal{A}_{\mathrm{det}}}
\newcommand{\Appgt}{\mathcal{A}_{\mathrm{gt}}}
\newcommand{\Appnovel}{\mathcal{A}_{\mathrm{det}} \setminus \mathcal{A}_{\mathrm{gt}}}
\newcommand{\TLSdet}{\mathcal{S}_{\mathrm{det}}}
\newcommand{\TLSgt}{\mathcal{S}_{\mathrm{gt}}}
\newcommand{\TLSnovel}{\mathcal{S}_{\mathrm{det}} \setminus \mathcal{S}_{\mathrm{gt}}}
\newcommand{\Rapp}{R_{\mathrm{app}}}
\newcommand{\Rappnovel}{R_{\mathrm{app}}^{\mathrm{novel}}}
\newcommand{\RTLS}{R_{\mathrm{TLS}}}
\newcommand{\RTLSnovel}{R_{\mathrm{TLS}}^{\mathrm{novel}}}

\subsubject{Metrics.} To measure capability of \tester in identifying vulnerable apps, we define \textit{Vulnerable App Detection Rate} ($\Rapp$) calculated as: 
    \[
        \Rapp = \frac{|\Appdet \cap \Appgt|}{|\Appgt|}
    \]
    where $\Appdet$ is the set of apps detected as vulnerable by \tester, and $\Appgt$ is the ground-truth set of vulnerable apps.

In addition, to measure the capability of \tester to go beyond the limitation of manual testing and identify new vulnerable apps, \textit{Novel Vulnerable App Discovery Rate} ($\Rappnovel$) is designed as: 
    \[
        \Rappnovel = \frac{|\Appnovel|}{|\Appdet|}
    \]
where $\Appnovel$ is the set of newly discovered vulnerable apps.

Similarly, we assess the detection effectiveness of \tester at the level of individual TLS flows using two metrics: (1) \textit{Vulnerable TLS Server Detection Rate} ($\RTLS$), defined as $\RTLS = \frac{|\TLSdet \cap \TLSgt|}{|\TLSgt|}$, where $\TLSdet$ is the set of vulnerable TLS servers detected by \tester and $\TLSgt$ is the ground-truth set; and (2) \textit{Novel Vulnerable TLS Server Discovery Rate} ($\RTLSnovel$), defined as $\RTLSnovel = \frac{|\TLSnovel|}{|\TLSdet|}$, where $\TLSnovel$ is the set of newly discovered vulnerable TLS servers. 

Notably, since the same TLS server (a distinct FQDN) may be contacted by multiple apps, and each app may implement TLS validation differently, we define a vulnerable TLS server in the context of each individual app when calculating these metrics.

\subsubject{Impact of LLM and Interaction Steps.} 
We first evaluate \tester under various combinations of interaction step budget ($N_{\mathrm{steps}}$) and the GUI interaction strategy ($S_{\mathrm{GUI}}$). Specifically, we consider step budgets of 10, 20, 50, and 100, and three GUI agent strategies: Random, General LLM, and Specialized LLM. 

\begin{table}
\centering
\footnotesize
\caption{End-to-end Detection Performance of \tester under Different Configurations}
\label{tab:tester_end2end}
\begin{threeparttable}
\begin{tabular}{lcccccc}
\toprule
\textbf{Strategy} & $N_{\mathrm{steps}}$\tnote{a} & $\Rapp$ & $\Rappnovel$ & $\RTLS$ & $\RTLSnovel$ \\
\midrule
Random         & 10  & 0.6957 & 0.0870 & 0.2930 & 0.2186 \\
Random         & 20  & 0.7391 & 0.1159 & 0.3147 & 0.2915 \\
Random         & 50  & 0.6812 & 0.1304 & 0.3473 & 0.5581 \\
Random         & 100 & 0.5797 & \textbf{0.1594} & 0.2589 & 0.4403 \\
\midrule
General LLM    & 10  & 0.7246 & 0.1014 & 0.2992 & 0.2884 \\
General LLM    & 20  & 0.7681 & 0.1304 & 0.3287 & 0.4574 \\
General LLM    & 50  & 0.7971 & 0.1304 & 0.3256 & 0.5473 \\
General LLM    & 100 & \textbf{0.8696} & 0.1449 & 0.4078 & \textbf{0.7302} \\
\midrule
Specialized LLM & 10  & 0.6812 & 0.0725 & 0.2791 & 0.2775 \\
Specialized LLM & 20  & 0.6812 & 0.0725 & 0.2450 & 0.3132 \\
Specialized LLM & 50  & 0.8116 & 0.1304 & \textbf{0.4186} & 0.4357 \\
Specialized LLM & 100 & 0.8261 & 0.0725 & 0.4000 & 0.4388 \\
\bottomrule
\end{tabular}
\begin{tablenotes}
\item [a] $N_{\mathrm{steps}}$ denotes the maximum number of GUI interaction steps per app.
\item [b] The interaction wait time ($T_{\mathrm{wait}}$) is set to 4 seconds for all experiments unless otherwise specified.
\item [c] The MitM retest policy ($P_{\mathrm{MitM}}$) is configured as ``always retest''; i.e., the MitM test is performed on all TLS flows.
\end{tablenotes}
\end{threeparttable}
\end{table}

Table~\ref{tab:tester_end2end} presents the end-to-end detection performance of \tester under these different configurations. The results indicate that both LLM-based strategies significantly outperform the Random baseline, and the highest detection rates are often achieved at a moderate budget of $N_{steps}$ = 50, with further steps yielding little to no improvement, suggesting that most apps are sufficiently explored within this limit.

\subsubject{Impact of the MitM Retest Policy.} Fixing the interaction step budget at 50, we then evaluate the impact of different MitM retest policies on the end-to-end performance of \tester. Specifically, we compare aforementioned three policies: (1) always retest, (2) always skip previously tested TLS servers, and (3) skip only if a previous test for that TLS server revealed a MitM vulnerability. Across these experiments we keep the GUI interaction strategy as Specialized LLM with a wait time of 4 seconds.

Table~\ref{tab:tester_mitm_policy} presents the detection performance of \tester under the three MitM retest policies. The results show that the "Always Skip" policy achieves the highest vulnerable app detection rate ($\Rapp$) and vulnerable TLS server detection rate ($\RTLS$). 
This suggests that redundant retesting of previously tested destinations does not improve detection effectiveness and may unnecessarily increase analysis time. Based on these findings, we adopt the "Always Skip" policy as the default configuration for subsequent experiments, as it provides an optimal balance between detection coverage and efficiency. 

\begin{table}
\centering
\footnotesize
\caption{End-to-End Detection Performance of \tester under Different MitM Retest Policies}
\label{tab:tester_mitm_policy}
\begin{threeparttable}
\begin{tabular}{lcccc}
\toprule
\textbf{MitM Retest Policy} & $\Rapp$ & $\Rappnovel$ & $\RTLS$ & $\RTLSnovel$ \\
\midrule
Always Retest & 0.8116 & 0.1304 & 0.4186 & 0.4357 \\  
Always Skip & \textbf{0.8986} & \textbf{0.1449} & \textbf{0.4279} & 0.2946 \\  
Skip if Vulnerable & 0.8261 & 0.1159 & 0.4093 & \textbf{0.5752} \\ 
\bottomrule
\end{tabular}
\begin{tablenotes}
\item The interaction step budget ($N_{\mathrm{steps}}$) is set to 50 and the GUI interaction strategy is Specialized LLM with a wait time of 4 seconds.

\end{tablenotes}
\end{threeparttable}
\end{table}

\subsection{Deployment of \tester}
\label{subsec:tester_deployment}
With \tester extensively evaluated, we proceed to deploy it for large-scale scanning of Android applications to detect TLS MitM vulnerabilities (TMVs). Below we elaborate the deployment setup and provide an overview of the deployment result.

\subject{Deployment Dataset.} We target two major datasets for large-scale deployment: (1) 20,000 apps from the official Google Play market, and (2) 20,000 apps from AppChina, a leading third-party app store serving mainland China. For both sources, apps are selected based on popularity, as approximated by download volume. This focus on popular apps is motivated by their widespread use and the likelihood that they handle sensitive user data, making potential TLS MitM vulnerabilities particularly impactful.

For Google Play, APKs are sourced from AndroZoo, a comprehensive and regularly updated Android app archive. For AppChina, APKs are downloaded directly from the store. In both cases, we select the latest available version of each app as of March 2025. After deduplication, the combined deployment dataset comprises 39,876 distinct apps and 40,118 unique APK files.

\subject{Deployment Configuration.} We configure \tester using the parameter settings that achieved the best trade-off between efficiency and effectiveness in our evaluation. Specifically, the GUI interaction strategy is set to Random, with an interaction wait time of 4 seconds. The MitM retest policy is set to ``Always Skip'' as this configuration skip previously tested TLS server. Then, each app is tested three times, each against one of the three different MitM tests. 

\subject{Deployment Results.} We applied \tester to the entire dataset of 40,118 APKs (corresponding to 39,876 unique app identifiers) in April, utilizing 8 concurrent Android emulators. The complete scan required 8 days to finish. In total, \tester successfully analyzed 37,349 apps (19,837 from AppChina and 17,512 from AndroZoo), with the left 2,769 ones failing the detection mostly due to startup timeouts, which occurred when incomplete application packages (e.g., split APKs missing native libraries) crashed upon launch or when apps failed to load essential network resources through our MitM proxy. The scan identified 8,374 apps exhibiting at least one TLS MitM vulnerability, corresponding to 5,881 unique vulnerable FQDNs. The average analysis time per app was 144.75 seconds, demonstrating the scalability and efficiency of \tester for real-world, market-scale deployment.

Next in Section~\ref{sec:tls_analysis}, we will try to extensively understand these detected TMVs.

\section{Understanding TLS MitM Vulnerabilities}
\label{sec:tls_analysis}
Using \tester, we successfully analyzed 37,349 apps and identified 8,374 apps vulnerable to one or more TLS Man-in-the-Middle vulnerabilities (TMVs). In this section, we systematically examine the prevalence and criticality of these TMVs, analyze their correlation with various app attributes, and investigate their longitudinal evolution in Android apps. Our goal is to provide a comprehensive understanding of the scope, impact, and persistence of TMVs within the Android ecosystem.

It is important to note that, compared to \textit{MitM-T1} and \textit{MitM-T2} (defined in Section~\ref{subsec:tester_design}), certificate pinning vulnerabilities (\textit{MitM-T3}) require a higher and less practical level of attack capability, specifically the installation of a root certificate on the Android device. Therefore, unless otherwise specified, we exclude certificate pinning-related TMVs from all aggregate measurement results involving TMVs. Instead, a dedicated analysis of certificate pinning vulnerabilities is provided in Section~\ref{subsec:analysis_cert_pin}.

\begin{table}  
    \centering  
    \caption{Fractions of apps and TLS flows that are vulnerable to TLS MitM vulnerabilities (TMVs).}  
    \label{tab:tmv_entity_fractions}  
    \scriptsize 
    \begin{threeparttable}  
        \scriptsize

        \begin{tabular}{@{}c@{}}
        \resizebox{\columnwidth}{!}{%
        \begin{tabular}{lccc}  
        \toprule  
        \textbf{Entity} & \textbf{AppChina} & \textbf{Androzoo} & \textbf{Both} \\  
        \midrule  
         Apps        & 7.82K/20K (\textbf{39.40\%}) & 558/18K (\textbf{3.19\%}) & 8.37K/37K (\textbf{22.42\%}) \\ 
        Flows       & 80K/800K (\textbf{9.94\%}) & 6.43K/839K (\textbf{0.77\%}) & 86K/1.64M (\textbf{5.25\%}) \\  
        FQDNs       & 5.04K/29K (\textbf{17.11\%}) & 919/20K (\textbf{4.69\%}) & 5.88K/48K (\textbf{12.16\%}) \\  
        A-FQDNs   & 30K/155K (\textbf{19.42\%}) & 1.61K/131K (\textbf{1.23\%}) & 32K/286K (\textbf{11.08\%}) \\  

        \bottomrule  
        \end{tabular}}
        \end{tabular}
        
        \begin{tablenotes}  
            \item \textit{Abbreviations:} Flows = TLS Flows; FQDNs = TLS Fully Qualified Domain Names; A-FQDNs = App-specific FQDNs.  
        \end{tablenotes}  
    \end{threeparttable}  
\end{table}

\subsection{How Prevalent are TMVs in Android Apps?}
\label{subsec:analysis_prevalence}
To assess prevalence of TMVs in Android apps, we begin by quantifying the extent to which TMV-related entities (apps and TLS traffic) contribute to the overall detection dataset. We further explore whether TMVs represent a widespread security issue across apps and TLS traffic varying in characteristics. This is achieved by analyzing the distribution patterns of TMVs with respect to app attributes (e.g., category, popularity) and TLS attributes (e.g., TLS versions). We also examine per-app TMV prevalence by measuring the proportion of TLS flows within a vulnerable app that are subject to TMVs. The following presents our results, accompanied by key summaries and discussions.

\subject{Scale and Fraction of Vulnerable Apps/TLS Traffic.} We first compute various fraction values to characterize the extent to which TMV-associated entities account for all entities of the same type under analysis. Four entity types of varying granularity are considered: TLS flows, app-specific TLS FQDNs (fully qualified domain names), TLS FQDNs, and apps. Notably, when counting TLS flows, each flow is uniquely identified by the tuple of app, TLS FQDN, and timestamp.

As presented in Table~\ref{tab:tmv_entity_fractions}, we summarize the fractions of TMV-related entities across the datasets. The table reports, for each entity type and dataset, both the absolute counts and the corresponding percentages of entities affected by TMVs, thus providing a clear overview of the scale and pervasiveness of TMVs in the Android app ecosystem. As we can see, a substantial portion of the ecosystem is affected, with 22.42\% of all unique apps (8,374 in total) exhibiting at least one TMVs. However, there is a stark difference in vulnerability prevalence between the two sources. The AppChina dataset shows a significantly higher rate of vulnerable apps (39.40\%) compared to the Androzoo dataset (3.19\%). This disparity holds true for all other metrics, such as vulnerable TLS flows (9.94\% for AppChina vs. 0.77\% for Androzoo). Furthermore, it is noteworthy that the fraction of vulnerable apps is considerably higher than the fraction of vulnerable flows (22.42\% vs. 5.25\% overall), suggesting that while many apps contain vulnerable code, these vulnerabilities may affect a small subset of their total TLS traffic.

\begin{figure}[t!]
    \centering
    \begin{subfigure}[b]{0.49\columnwidth}
        \centering
        \includegraphics[width=\textwidth]{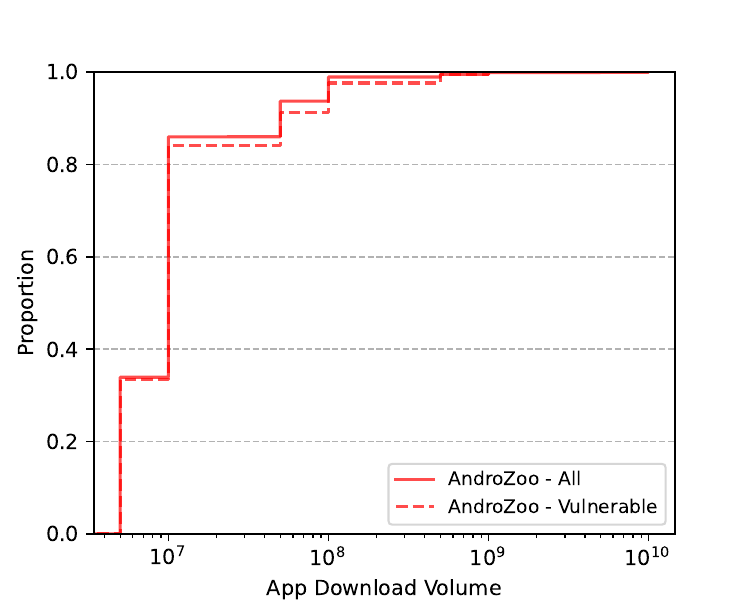}
        \caption{App download volume}
        \label{fig:cdf_downloads}
    \end{subfigure}
    \hfill
    \begin{subfigure}[b]{0.49\columnwidth}
        \centering
        \includegraphics[width=\textwidth]{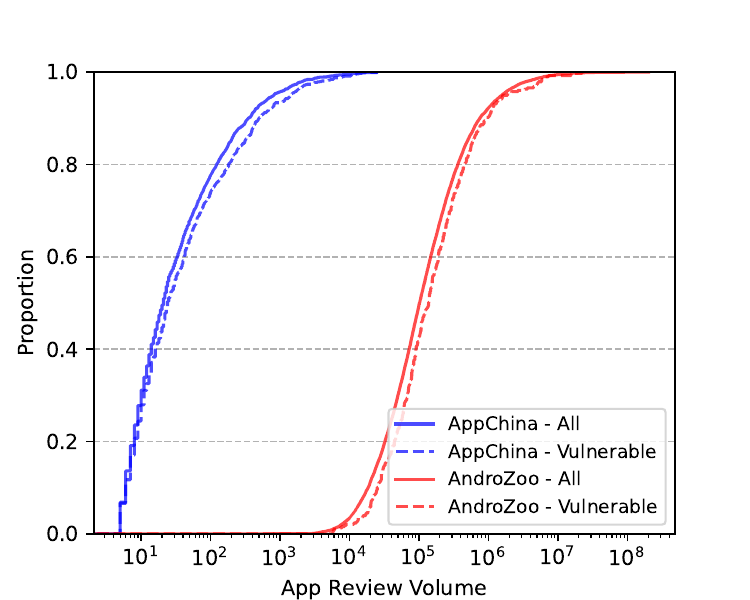}
        \caption{App review volume}
        \label{fig:cdf_reviews}
    \end{subfigure}
    
    \caption{Cumulative distribution functions (CDFs) of vulnerable apps (TMVs) and all apps over log-scaled popularity metrics. Subfigure~\ref{fig:cdf_downloads} shows the distribution by download volume, while Subfigure~\ref{fig:cdf_reviews} shows it by review volume. Notably, as the download volume is not available for the AppChina dataset, Subfigure~\ref{fig:cdf_downloads} depicts only AndroZoo. 
    }
    \label{fig:tmv_cdf_popularity}
\end{figure}

\subject{Correlation between TMVs and App/TLS Characteristics.} To further understand the generality of TMVs, we analyze their distribution across key app characteristics, specifically app popularity and category. Regarding distribution across app popularity, we adopt two metrics: app download volume and review volume. We then calculate the cumulative distribution functions (CDFs) of vulnerable apps as a function of these log-scaled metrics, comparing them directly with all apps in the detection dataset. As shown in Figure~\ref{fig:tmv_cdf_popularity}, the CDF curves for vulnerable apps (dashed lines) almost perfectly overlap with the curves for all apps (solid lines) across both the AppChina and Androzoo datasets. This visual alignment suggests that TMVs are not concentrated in apps of a particular popularity tier but are instead distributed evenly across the entire popularity spectrum. This observation is quantitatively supported by the point-biserial correlation coefficients ($r_{pb}$). The correlation between app vulnerability and log-scaled download volume (Androzoo) is only -0.0012, and it is similarly negligible for log-scaled review volume, with rpb values of 0.0005 (Androzoo) and 0.0366 (AppChina). All coefficients indicate a very weak, insignificant correlation.

In addition, to further examine the relationship between TMVs and app categories, we analyze the distribution of both vulnerable and all apps across the top app categories by app count, as summarized in Table~\ref{tab:category_distributions}. To quantitatively assess the similarity between the category distributions of vulnerable and all apps, we employ the Jensen-Shannon Divergence (JSD), a widely used metric for comparing similarity between two probability distributions: The closer the JSD value is to zero, the greater similarity the two distributions are of. The resulting JSD values are 0.0499 for AppChina and 0.2433 for Androzoo. The low JSD for AppChina indicates a high similarity between the category distributions of vulnerable and all apps. The value for Androzoo, while higher, also suggests a substantial overlap, confirming that vulnerable apps are not concentrated in a few specific categories.

Our results reveal that TMVs are not confined to less popular or niche categories; instead, they are prevalent across a broad spectrum of app popularity and categories. This strongly indicates that TMVs are a widespread issue, affecting both highly popular and less prominent apps, as well as a diverse range of app types.

\begin{table}
    \centering
    \caption{Distribution of vulnerable and all apps among top 10 app categories.}
    \label{tab:category_distributions}
    \begin{threeparttable}
        \begin{tabular*}{\linewidth}{@{\extracolsep{\fill}}cc}
            \begin{subtable}{.48\linewidth}
                \centering
                \caption{AppChina}
                \label{tab:top10_appchina}
                \scriptsize
                \setlength{\tabcolsep}{4pt}
                \begin{tabular}{lcc}
                \toprule
                \textbf{Category} & \textbf{Vulnerable} & \textbf{All} \\
                \midrule
                Productivity & 27.28\% & 27.87\% \\
                Lifestyle & 23.94\% & 23.90\% \\
                Tools & 12.70\% & 14.98\% \\
                Shopping & 11.72\% & 8.96\% \\
                Social & 6.78\% & 5.84\% \\
                Photography & 3.13\% & 4.09\% \\
                Finance & 2.71\% & 3.55\% \\
                Books & 4.13\% & 3.52\% \\
                Media & 3.17\% & 2.81\% \\
                News & 2.24\% & 2.09\% \\
                \bottomrule
                \end{tabular}
            \end{subtable}
        &
            \begin{subtable}{.48\linewidth}
                \centering
                \caption{Androzoo}
                \label{tab:top10_androzoo}
                \scriptsize
                \setlength{\tabcolsep}{4pt}
                \begin{tabular}{lcc}
                \toprule
                \textbf{Category} & \textbf{Vulnerable} & \textbf{All} \\
                \midrule
                Simulation & 5.20\% & 11.94\% \\
                Action & 3.23\% & 8.79\% \\
                Puzzle & 3.58\% & 6.55\% \\
                Tools & 6.09\% & 6.30\% \\
                Entertainment & 4.48\% & 3.58\% \\
                Photography & 1.97\% & 2.78\% \\
                Education & 1.08\% & 2.58\% \\
                Finance & 2.15\% & 2.56\% \\
                Racing & 0.36\% & 2.50\% \\
                Shopping & 4.66\% & 2.22\% \\
                \bottomrule
                \end{tabular}
            \end{subtable}
        \end{tabular*}

        \begin{tablenotes}
        \item \textit{Note:} Each value represents the proportion of either vulnerable or all apps within the specified app category. This comparison highlights whether certain categories are disproportionately affected by TMVs.
        \end{tablenotes}
    \end{threeparttable}
\end{table}

\begin{figure}
    \centering
    \includegraphics[width=0.48\textwidth]{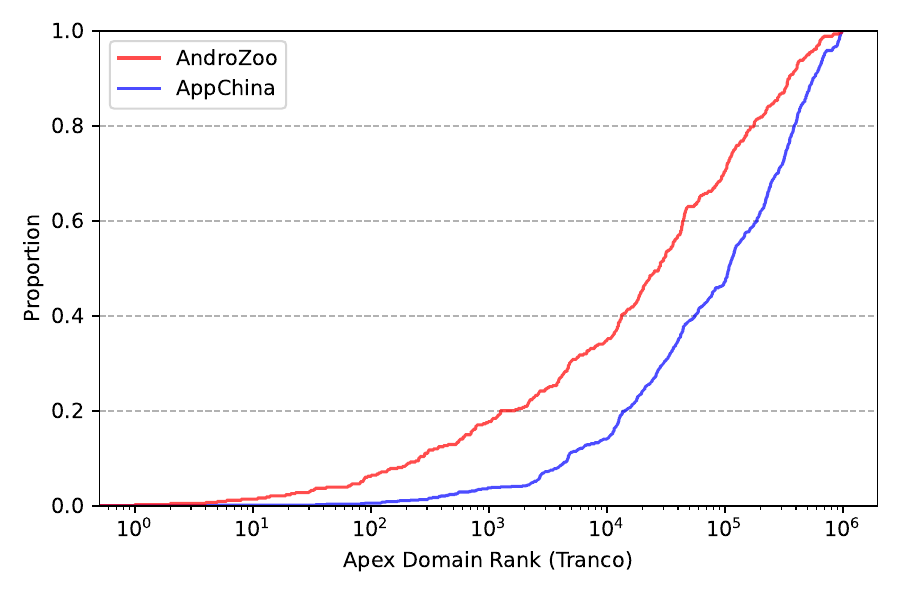}
    \caption{Cumulative distribution of vulnerable TLS FQDNs over log-scaled apex domain popularity (Tranco ranking).}
    \label{fig:tmv_cdf_fqdn_popularity}
\end{figure}

We further examined the correlation between TMVs and key TLS characteristics, specifically TLS version and the popularity of TLS FQDNs. Table~\ref{tab:dist_tmv_over_tls_attrs} summarizes the distribution of vulnerable TLS flows across different TLS versions. As shown, TMVs predominantly occur in TLS 1.3 flows (78.98\%), although a significant portion is also found in TLS 1.2 flows (21.02\%). 
To assess the prevalence of TMVs with respect to the popularity of TLS FQDNs, we approximate FQDN popularity using the Tranco top 1 million domain ranking of their apex domains~\cite{LePochat2019}. A TLS FQDN is considered as vulnerable if any of its associated TLS flows exhibit TMVs. Figure~\ref{fig:tmv_cdf_fqdn_popularity} illustrates the cumulative distribution of vulnerable TLS FQDNs over log-scaled apex domain popularity. As observed, vulnerable FQDNs are distributed across the entire popularity spectrum, rather than being concentrated in either highly popular or obscure domains. For instance, in the Androzoo dataset, approximately 20\% of vulnerable FQDNs are linked to apex domains within the top 10,000 ranks, while the remaining 80\% are associated with less popular, long-tail domains. A similar trend is observed in the AppChina dataset.

\begin{table}  
    \centering  
    \caption{Distribution of vulnerable and all TLS flows across TLS versions and network protocols.}  
    \label{tab:dist_tmv_over_tls_attrs}  
    \begin{threeparttable}  
        \scriptsize  
        \begin{tabular}{lcccc}  
        \toprule  
        \textbf{Category} & \multicolumn{2}{c}{\textbf{TLS Version}} & \multicolumn{2}{c}{\textbf{Transport Protocol}} \\  
        \cmidrule(lr){2-3} \cmidrule(lr){4-5}  
         & TLS 1.2 & TLS 1.3 & TCP & UDP  \\  
        \midrule  
        Vulnerable\tnote{a} & 21.02\% & 78.98\% & 100.00\% & 0.00\% \\  
        All & - & - & 99.78\% & 0.22\% \\  
        \bottomrule  
        \end{tabular}  
        \begin{tablenotes}  
            \item [a] Vulnerable denotes TLS flows subject to TMVs.  
        \end{tablenotes}  
    \end{threeparttable}  
\end{table}

\subsection{How Damaging are TMVs in Android Apps?}
\label{subsec:analysis_criticality}
Having established that TMVs are a prevalent issue, widely distributed across diverse app characteristics and TLS attributes, we now turn to evaluating their potentially damaging impact, i.e., the criticality of TMVs. The criticality of a TMV depends largely on the nature of the affected TLS flows. For example, if a TMV compromises flows transmitting sensitive information such as login credentials or personal data, the consequences can be severe, potentially leading to significant privacy breaches or unauthorized account access. Conversely, if the vulnerable flows are limited to transmitting non-sensitive data, such as app performance metrics, the potential damage is considerably lower.

To systematically assess the impact of TMVs, we analyze both the proportion of an app's TLS traffic that is vulnerable and the types of data or functionalities exposed. This includes identifying whether the affected flows are associated with critical operations, such as authentication, financial transactions, or the transmission of personally identifiable information. Through case studies of popular apps and aggregate statistical measurements, we provide a comprehensive evaluation of the real-world risks posed by TMVs in Android applications.

\subject{Ratio of Vulnerable TLS Traffic per App.} To assess the potential security impact of TMVs at the app level, we analyze the ratio of vulnerable TLS flows within each affected app. Specifically, for every app identified as vulnerable, we calculate the proportion of its total TLS flows that are subject to TMVs. This metric offers valuable insight into whether TMVs compromise only a minor subset of an app's traffic or represent a systemic issue affecting a substantial portion of its encrypted communications. A higher value of this ratio indicates a greater potential for damage, as a larger share of the app's TLS traffic is exposed to interception and manipulation.

\begin{figure}
    \centering
     \includegraphics[width=0.48\textwidth]{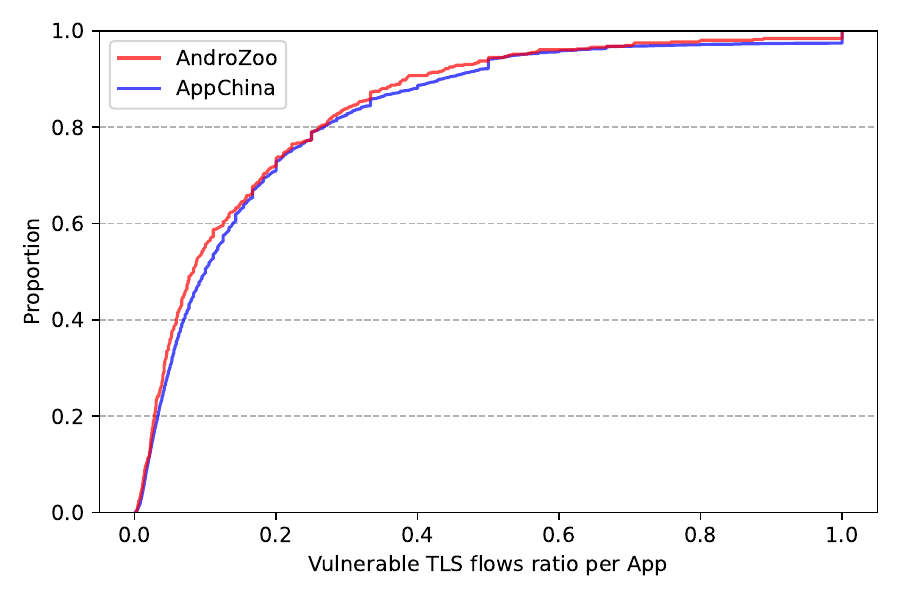}
    \caption{Cumulative distribution of the ratio of vulnerable TLS flows per app in AppChina and Androzoo datasets.}
    \label{fig:tmv_flow_ratio_per_app}
\end{figure}

Figure~\ref{fig:tmv_flow_ratio_per_app} illustrates the cumulative distribution function (CDF) of the ratio of vulnerable TLS flows per app for both the AppChina and Androzoo datasets. The results reveal that a significant fraction of vulnerable apps have a considerable proportion of their TLS traffic exposed to TMVs. For example, 50\% of vulnerable apps in AppChina have at least 10\% of their TLS flows affected, while in Androzoo, the corresponding figure is 8.33\%. Notably, a non-trivial number of apps exhibit ratios approaching 1, indicating that nearly all their TLS traffic is susceptible to TMVs. These findings highlight that, for many apps, TMVs are not isolated incidents but rather pervasive issues that can compromise the majority of their encrypted communications.

To complement this analysis, Table~\ref{tab:tmv_flow_ratio_stats} provides a summary of key statistics for the per-app vulnerable flow ratios, including the median, mean, and the percentage of apps with more than 50\% of their TLS flows vulnerable. These results reinforce the observation that, for a significant subset of apps, TMVs are not isolated incidents but rather affect a substantial portion of their encrypted communications, thereby amplifying the associated security and privacy risks.

\begin{table}
    \centering
    \caption{Statistics of per-app ratio of vulnerable TLS flows.}
    \label{tab:tmv_flow_ratio_stats}
    \footnotesize
    \begin{threeparttable}
        \scriptsize
        \begin{tabular}{lccc}
        \toprule
        \textbf{Dataset} & \textbf{Median Ratio} & \textbf{Mean Ratio} & \textbf{\% Apps with Ratio $>$ 0.5} \\
        \midrule
        AppChina & 0.1000 & 0.1726 & 5.90\% \\
        Androzoo & 0.0833 & 0.1596 & 5.56\% \\
        \bottomrule
        \end{tabular}
        \begin{tablenotes}
            \item \textit{Note:} Ratio is defined as the number of vulnerable TLS flows divided by the total number of TLS flows per app.
        \end{tablenotes}
    \end{threeparttable}
\end{table}

The observation that not all TLS flows within a vulnerable app are themselves vulnerable suggests that both secure and insecure TLS validation processes can coexist within a single app. This highlights the complexity of TMVs and motivates a deeper investigation into their root causes, which we elaborate on in Section~\ref{sec:tls_causality}.

\subject{Confirming and Profiling Critical but Vulnerable TLS Flows via Manual Labeling.} Although a high ratio of vulnerable TLS flows can indicate a great potential of risks, a more direct evidence is the extent to which \textit{critical} TLS flows per app are subject to TMVs.

To assess this, we conducted a detailed case study on a subset of 100 highly popular vulnerable apps (50 from AppChina and 50 from Androzoo). Our researchers manually interacted with each app, exercising key functionalities such as login, registration, shopping and payment. During these interactions, we performed MitM tests and labeled each TLS flow according to the functionality being exercised at the time of transmission.

We defined five categories of TLS flows based on their functional context and criticality:  
\begin{enumerate}  
    \item \textit{Authentication}: Flows related to user login, registration, or session management.  
    \item \textit{Financial Transaction}: Flows involving payment, in-app purchases, or financial data exchange.  
    \item \textit{Content Delivery}: Flows delivering app content, such as media, news, or product information.  
    \item \textit{Telemetry/Analytics}: Flows related to app analytics, usage statistics, or non-sensitive background data.  
    \item \textit{Executable Code}: Flows involving dynamic libraries, application updates, or patches.
\end{enumerate}

Table~\ref{tab:critical_flow_categories} summarizes the proportion of vulnerable TLS flows in each category across the studied apps. Notably, a significant fraction of critical flows—especially those related to authentication and executable code—were found to be vulnerable to TMVs.

\begin{table}  
    \centering  
    \caption{Proportion of Vulnerable TLS Flows by Functional Category (Manual Labeling of 100 Popular Apps)}  
    \label{tab:critical_flow_categories}  
    \footnotesize  
    \begin{threeparttable}  
        \scriptsize  
        \begin{tabular}{lcc}  
        \toprule  
        \textbf{Category} & \textbf{Vulnerable Flows (\%)} & \textbf{Vulnerable Apps(\%)} \\  
        \midrule  
        Content Delivery       & 61.28 & 56.00 \\
        Telemetry/Analytics    & 27.70 & 61.00 \\
        Executable Code        & 6.19 & 27.00 \\
        Authentication         & 4.06 & 39.00 \\
        Financial Transactions & 0.75 & 13.00 \\
        \bottomrule  
        \end{tabular}  
        \begin{tablenotes}  
            \item \textit{Note:} The first column shows the percentage of TLS flows in each category that are vulnerable; the second column shows the percentage of apps in which at least one TLS flow of the category is vulnerable.  
        \end{tablenotes}  
    \end{threeparttable}  
\end{table}

\begin{figure}
    \centering
    \includegraphics[width=0.48\textwidth]{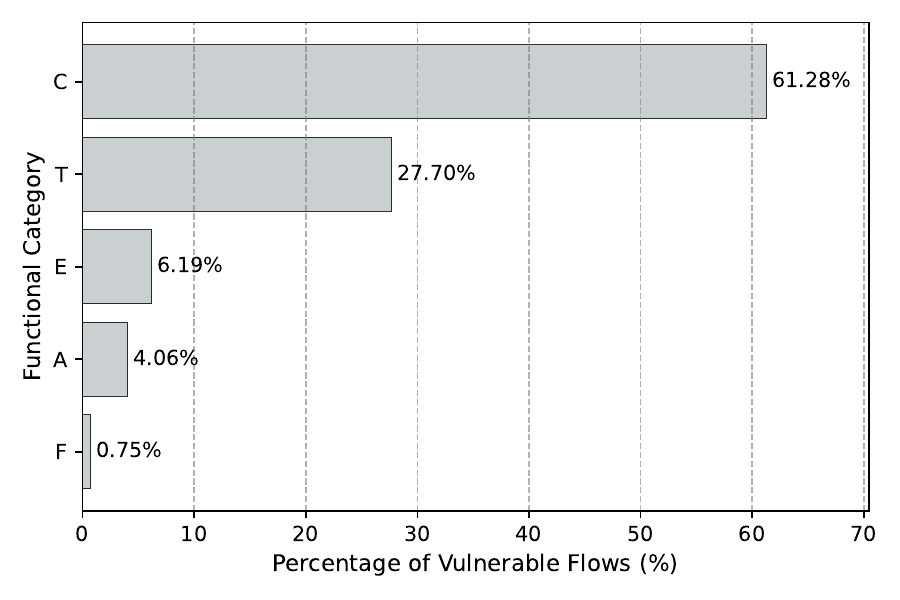}
    \caption{Distribution of vulnerable TLS flows by functional category across 100 popular apps: (C) Content Delivery, (T) Telemetry/Analytics, (E) Executable Code, (A) Authentication, and (F) Financial Transaction.}
    \label{fig:critical_flow_vuln_ratio}
\end{figure}

Figure~\ref{fig:critical_flow_vuln_ratio} further visualizes the distribution of vulnerable flows by category. The results reveal that TMVs are not limited to non-critical background traffic; rather, they frequently affect flows carrying executable code, sensitive user credentials, and financial information. For example, 6.19\% of executable code flows, 4.06\% of authentication flows, and 0.75\% of financial transaction flows were found to be vulnerable, highlighting the real-world risks of arbitrary code execution, credential theft, and unauthorized transactions.

\subsubject{Observations.} Our manual labeling and analysis demonstrate that a substantial portion of TMVs directly impact critical app functionalities. This underscores the severity of the threat: executable code, users' authentication credentials, and financial transactions are frequently exposed to interception risks due to improper TLS validation. These findings highlight the urgent need for app developers to prioritize secure TLS implementations, especially for flows handling sensitive operations.

\subsection{How Do TMVs Evolve over Time in Android Apps?}
\label{subsec:analysis_evolution}

To better understand the persistence and remediation of TMVs, we conduct a longitudinal analysis of their evolution in Android apps. Ideally, we would like to examine how long TMVs persist in all affected apps, whether these vulnerabilities are promptly addressed in subsequent versions, and if new vulnerabilities are introduced over time. However, due to the limited availability of complete version histories for all apps, we focus on a representative subset for which sufficient historical data is available.

\subject{Methodology and Metrics.} We selected 100 apps with confirmed TLS Man-in-the-Middle (MITM) vulnerabilities and collected their historical versions from app markets such as Wandoujia and the AndroZoo database, covering the last 5 years. To ensure temporal granularity while avoiding redundancy, we sampled one APK per app for each non-overlapping three-month interval, resulting in a total of 1,720 APKs for in-depth analysis.

For each app, we define the following metrics:
\begin{itemize}
    \item \textit{Vulnerable Span (days):} The total of the three-month time slots. A time slot is counted during only when the respective APK version exhibits TMVs.
    \item \textit{App Lifespan (days):} The time span from the earliest to the latest available version.
    \item \textit{Vulnerable Span Ratio:} The ratio of the vulnerable span to the app lifespan, indicating the persistence of TMVs relative to the app's history.
    \item \textit{Remediation Delay (days):} For apps where TMVs are eventually fixed, the time elapsed between the first detection and the first non-vulnerable version.
    \item \textit{Reintroduction Events:} The number of times TMVs are reintroduced after being fixed in a previous version.
\end{itemize}

\subject{Results and Observations.} Table~\ref{tab:tmv_evolution_stats} summarizes key statistics of TMV evolution across the analyzed apps. The results show that TMVs are often long-lived: the median vulnerable span is 1384 days, and the median vulnerable span ratio is 92.10\%, indicating that vulnerabilities persist for a substantial portion of the app's lifecycle. Furthermore, only 13\% of apps ever remediated TMVs within the observed period, and among those, the median remediation delay is 330 days. Notably, in 39\% of cases, TMVs were reintroduced after being fixed, highlighting the risk of regression.

\begin{table}
    \centering
    \caption{Statistics of TMV evolution across historical app versions.}
    \label{tab:tmv_evolution_stats}
    \footnotesize
    \begin{threeparttable}
        \scriptsize
        \begin{tabular}{lcccc}
        \toprule
        \textbf{Metric} & \textbf{Median} & \textbf{Mean} & \textbf{Min} & \textbf{Max} \\
        \midrule
        Vulnerable Span (days) & 1384 & 1350.26 & 0 & 2037 \\
        App Lifespan (days) & 1901 & 1646.47 & 0 & 2067 \\
        Vulnerable Span Ratio (\%) & 92.10 & 82.89 & 0.00 & 100.00 \\
        Remediation Delay (days) & 330 & 486.49 & 62 & 1763 \\
        Reintroduction Events (count) & 0 & 0.61 & 0 & 4 \\
        \bottomrule
        \end{tabular}
        \begin{tablenotes}
            \item \textit{Note:} Metrics are computed over the set of 100 apps with sufficient historical data.
        \end{tablenotes}
    \end{threeparttable}
\end{table}

\begin{figure}
    \centering
    \includegraphics[width=0.45\textwidth]{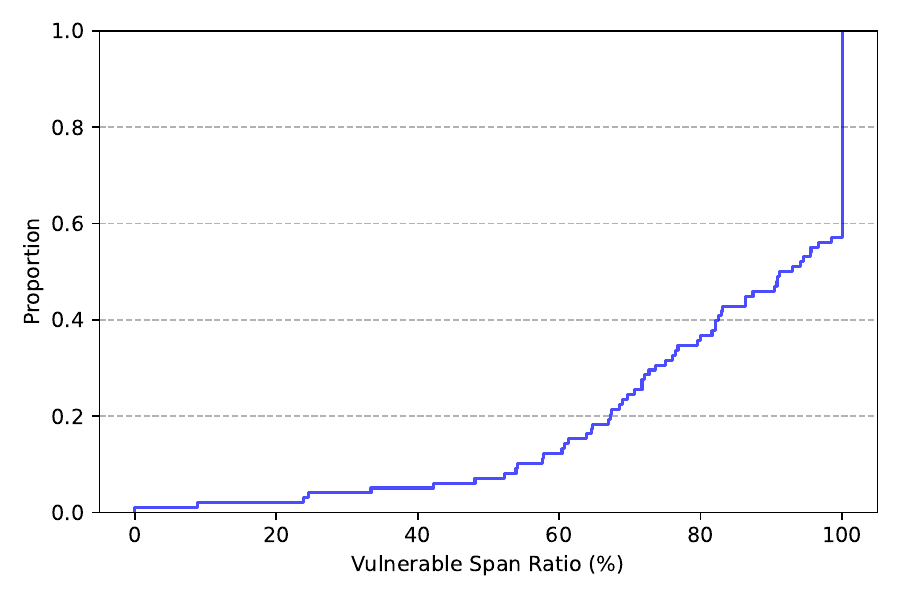}
    \caption{CDF of the ratio of vulnerable span to app lifespan for apps with historical TMV data.}
    \label{fig:tmv_vuln_span_cdf}
\end{figure}

Figure~\ref{fig:tmv_vuln_span_cdf} presents the cumulative distribution function (CDF) of the vulnerable span ratio across all studied apps. As shown, a significant fraction of apps have TMVs persisting for over 100\% of their observed lifespan, and only a small minority achieve timely remediation.

\subsubject{Observations:} Our longitudinal analysis reveals that TMVs are not only widespread but also highly persistent in Android apps. Most vulnerabilities remain unaddressed for the majority of the app's lifetime, and even when fixed, they are sometimes reintroduced in later versions. This underscores the need for continuous security monitoring and more robust development practices to prevent both the persistence and recurrence of TMVs throughout the app lifecycle.

\subsection{The Certificate Pinning Vulnerability}
\label{subsec:analysis_cert_pin}
When generating the above measurement results, we exclude the certificate pinning vulnerability (\textit{MitM-T3}). This is because it requires the attacker to directly install a root CA certificate on a target Android device, thus being much less realistic than the other two TMV types. Now, we present the detection results on certificate pinning, with a focus on the fraction of apps and TLS flows that are subject to this vulnerability.

\subject{Scale and Fraction of Vulnerable Apps/TLS Traffic.}
To quantify the prevalence of certificate pinning vulnerabilities, we analyze both the number and proportion of affected apps and TLS flows in our datasets. Table~\ref{tab:cert_pin_entity_fractions} summarizes the absolute counts and percentages of apps, TLS flows, TLS FQDNs, and app-FQDN pairs exhibiting certificate pinning vulnerabilities in both AppChina and Androzoo datasets.

\begin{table}  
    \centering  
    \caption{Fractions of apps and TLS flows vulnerable to TLS certificate pinning bypass.}  
    \label{tab:cert_pin_entity_fractions}  
    \footnotesize  
    \begin{threeparttable}  
        \scriptsize  
        \begin{tabular}{@{}c@{}}
        \resizebox{\columnwidth}{!}{%
        \begin{tabular}{lccc}  
        \toprule  
        \textbf{Entity} & \textbf{AppChina} & \textbf{Androzoo} & \textbf{Both} \\  
        \midrule  
         Apps        & 13K/19K (\textbf{65.77\%}) & 15K/18K (\textbf{84.56\%}) & 28K/37K (\textbf{74.84\%}) \\  
        Flows       & 1.12M/1.20M (\textbf{93.68\%}) & 828K/1.04M (\textbf{79.96\%}) & 1.95M/2.23M(\textbf{87.32\%}) \\ 
        FQDNs       & 26K/27K (\textbf{98.68\%}) & 21K/22K (\textbf{98.48\%}) & 47K/48K (\textbf{98.62\%}) \\  
        A-FQDNs   & 138K/145K(\textbf{94.73\%}) & 159K/160K(\textbf{99.12\%}) & 296K/305K (\textbf{97.03\%}) \\ 
        \bottomrule  
        \end{tabular}}
        \end{tabular}
        \begin{tablenotes}  
            \item \textit{Abbreviations:} Flows = TLS Flows; FQDNs = TLS Fully Qualified Domain Names; A-FQDNs = App-specific TLS Fully Qualified Domain Names.  
        \end{tablenotes}    
    \end{threeparttable}  
\end{table}

\begin{figure}
    \centering
    \begin{subfigure}[b]{0.49\columnwidth}
        \centering
        \includegraphics[width=\textwidth]{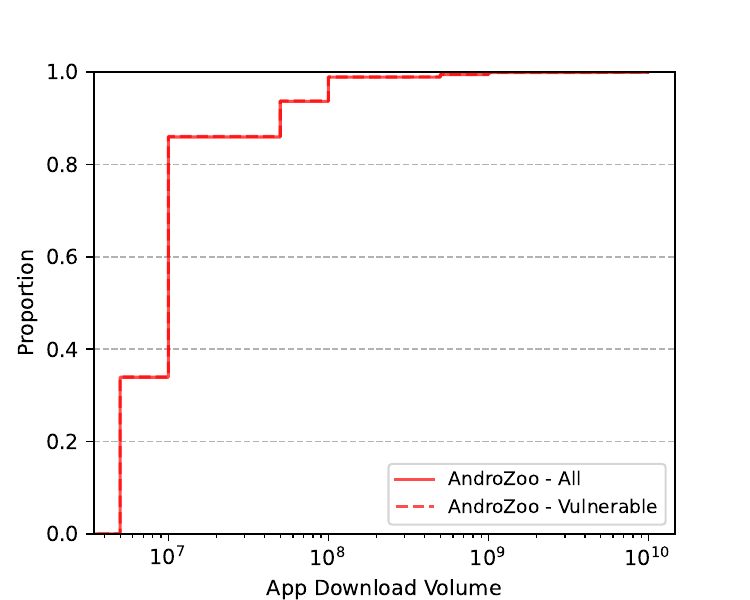}
        \caption{App download volume}
        \label{fig:cert_pin_cdf_downloads}
    \end{subfigure}
    \hfill
    \begin{subfigure}[b]{0.49\columnwidth}
        \centering
        \includegraphics[width=\textwidth]{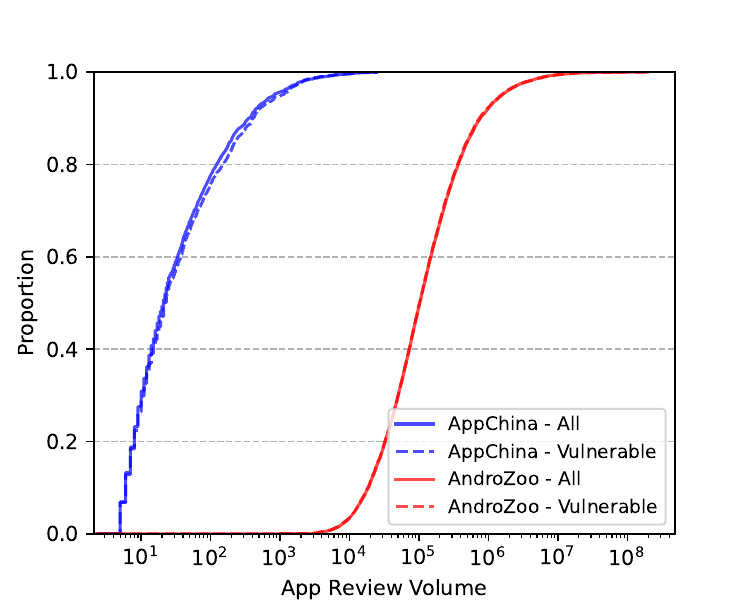}
        \caption{App review volume}
        \label{fig:cert_pin_cdf_reviews}
    \end{subfigure}
    \caption{Cumulative distribution of certificate pinning vulnerable apps and all apps over log-scaled popularity metrics. Subfigure (a) shows the distribution by download volume, while (b) shows it by review volume.}
    \label{fig:cert_pin_cdf_popularity}
\end{figure}

Figure~\ref{fig:cert_pin_cdf_popularity} visualizes the cumulative distribution of certificate pinning vulnerable apps over log-scaled popularity metrics, specifically app download and review volume, compared to all apps. This allows us to assess whether such vulnerabilities are concentrated among less popular apps or distributed across the popularity spectrum.

\subject{Observations.}
Our results indicate that certificate pinning vulnerabilities, despite being less practical,  are significantly more prevalent than the other two TMV types. Specifically, approximately 75\% of apps and over 98\% of FQDNs are affected, as shown in Table~\ref{tab:cert_pin_entity_fractions}. The distribution in Figure~\ref{fig:cert_pin_cdf_popularity} shows that these vulnerabilities are not limited to obscure or unpopular apps; a non-trivial fraction of popular apps also exhibit improper or bypassable certificate pinning. This suggests that, while less practical, certificate pinning vulnerabilities still pose a concerning risk, especially in high-profile apps where users may expect stronger security guarantees.

These findings highlight the importance of not only implementing certificate pinning but also ensuring its correct and robust deployment. Improper pinning implementations can create a false sense of security and expose users to sophisticated MitM attacks, particularly in scenarios where attackers can gain privileged device access.

\subsection{Summary}
Our comprehensive measurement study reveals that TLS man-in-the-middle vulnerabilities (TMVs) are both widespread and persistent in the Android app ecosystem. TMVs affect a substantial fraction of apps and TLS flows, with vulnerabilities distributed across all levels of app popularity and categories, as well as across different TLS versions and domain popularities. Critically, TMVs are not confined to non-essential traffic; they frequently compromise sensitive operations such as authentication, financial transactions, and personal data transmission. Manual analysis of popular apps confirms that a significant portion of \textit{critical} flows are exposed to TMVs, amplifying the real-world risks to user security and privacy.

Longitudinal analysis further demonstrates that TMVs often persist for extended periods, with many vulnerabilities remaining unaddressed throughout much of an app's lifecycle. Even when remediated, TMVs are sometimes reintroduced in later versions, highlighting the challenges of sustainable vulnerability management. Additionally, certificate pinning vulnerabilities are found to be even more prevalent, affecting a large share of apps and TLS flows—including many popular apps—underscoring the need for correct and robust pinning implementations.

These findings underscore the urgent need for scalable, systematic approaches to vulnerability analysis and disclosure. In the following sections, we conduct a root cause analysis of TMVs (Section~\ref{sec:tls_causality}) and present a framework for scalable responsible disclosure (Section~\ref{sec:disclosure}), leveraging large language models (LLMs) to empower both processes.

\section{Root Cause Analysis of TMVs}
\label{sec:tls_causality}
Building on the analysis of TLS validation vulnerabilities (TMVs) with regards to prevalence and criticality, we now turn to investigating their underlying causes. This section presents our origin and root cause analysis (ORCA) methodology for TMVs along with analysis results. We first motivate the need for ORCA by highlighting the observed diversity in TMV manifestations across flows and domains. Next, we introduce our tool, \analyzer, and describe the instrumentation techniques employed to precisely locate vulnerable code within applications. We then detail the process of defining a taxonomy of TLS validation errors and present our LLM-based approach for categorizing these errors at scale. Finally, we analyze the distribution of error categories and responsible parties, offering insights into common root causes and informing future mitigation efforts.

\begin{figure}
    \centering
    \includegraphics[width=0.48\textwidth]{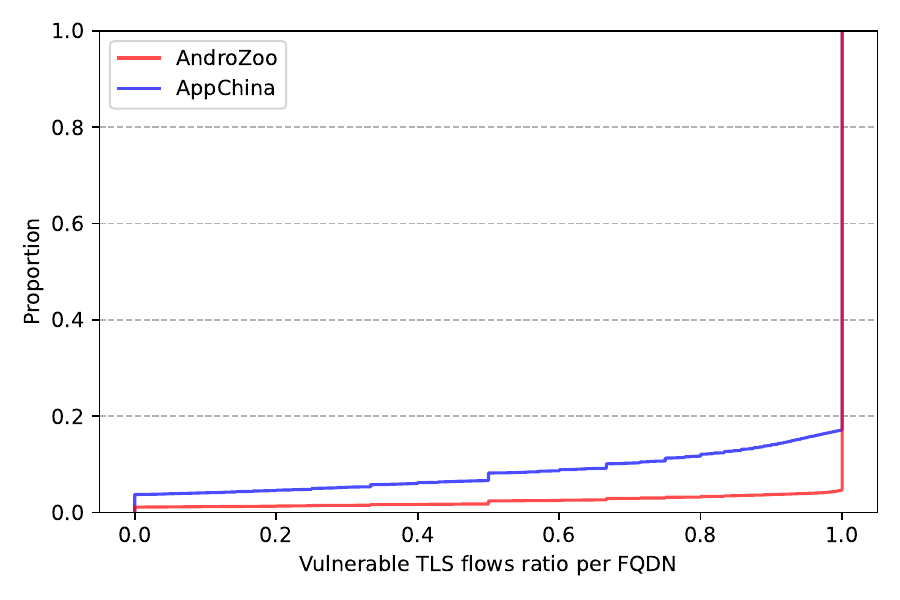} %
    \caption{Cumulative distribution function (CDF) of a vulnerable TLS FQDN over the ratio of vulnerable TLS flows towards it.}
    \label{fig:tls_flow_vul_ratio_cdf}
\end{figure}

\subject{Motivation.}
Several factors motivate a deeper investigation into the root causes of TLS validation vulnerabilities (TMVs). As discussed in Section~\ref{subsec:analysis_criticality}, the majority of vulnerable applications do not exhibit TMVs across all their TLS flows. In fact, 94.12\% of cases show that at least half of the flows are free from such vulnerabilities. Furthermore, Figure~\ref{fig:tls_flow_vul_ratio_cdf} presents the cumulative distribution function (CDF) of vulnerable TLS FQDNs with respect to the proportion of their associated flows that are vulnerable. This demonstrates that, while the majority of FQDNs have all their TLS traffic affected by TMVs, some FQDNs still have a portion of their TLS flows free from such vulnerabilities.

These observations highlight the need to understand why certain flows are vulnerable while others are not, even within the same application or FQDN. To provide actionable explanations and to inform effective vulnerability disclosure and remediation, we conduct a comprehensive root cause analysis of TMVs. Our analysis is designed to achieve two primary objectives: (1) to precisely locate and categorize the vulnerable code responsible for each TMV, and (2) to attribute each vulnerability to the appropriate responsible party, such as the app developer or a third-party library provider. Next, we move to present the design of our tool for \textbf{O}rigin and \textbf{R}oot \textbf{C}ause \textbf{A}nalysis tool, namely, \analyzer.

\subsection{{\analyzer}: Origin and Root Cause Analysis for TLS MitM Vulnerabilities}

To address the challenge of scalable and reliable ORCA (origin and root cause analysis) for TLS validation vulnerabilities (TMVs), we developed \analyzer, an automated toolchain designed to systematically identify and categorize the underlying causes of TMVs in Android applications as well as attributing TMVs to the responsible parties.

\subject{Limitations of Prior RCA Approaches.}
Existing ORCA efforts have several notable limitations. First, they are often manual and labor-intensive, relying on code inspection or small-scale static analysis, which does not scale to the thousands of apps and codebases as detected in this study. Second, prior analyses are typically restricted to a small set of applications or outdated datasets, limiting their coverage and resulting in findings that may not reflect the current vulnerability landscape. Third, previous work frequently employs coarse-grained or ad-hoc categories for vulnerable code, lacking the granularity needed to distinguish between different error patterns and their security implications. Finally, the rapid evolution of Android apps and third-party libraries means that manually curated ORCA results quickly become obsolete.

\subject{Overview of \analyzer.} \analyzer overcomes these limitations through automation and a modular design, comprising two primary components.
The first is the \texttt{Vulnerable Code Locator}, which leverages dynamic instrumentation to precisely identify the code locations responsible for TLS validation failures at runtime. By monitoring relevant classes and methods as they are loaded and executed, TMV-RCA achieves high coverage even in the presence of dynamic class loading and code obfuscation. The second is the \texttt{Vulnerable Code Classification} module, which, once vulnerable code snippets are extracted, employs a Large Language Model (LLM)-based classifier to assign fine-grained error categories according to a comprehensive taxonomy. This enables scalable, consistent, and up-to-date categorization of vulnerabilities across large datasets, capturing nuanced error patterns that manual approaches often miss.
By integrating these modules, \analyzer enables scalable, fine-grained root cause analysis for TMVs, delivering actionable insights for developers, security analysts, and the research community. 

In the following subsections, we detail each module of \analyzer in Section~\ref{subsec:rca_code_locator} and Section~\ref{subsec:rca_code_classify}, while presenting the results of our ORCA in Section~\ref{subsec:rca_result}.
\subsection{Vulnerable Code Locator of \analyzer}

\label{subsec:rca_code_locator}
As a core component of \analyzer, the Vulnerable Code Locator is responsible for dynamically instrumenting vulnerable Android applications to capture comprehensive contextual information about TLS validation events—such as the certificates being validated and their associated FQDNs. This detailed runtime logging enables precise correlation between specific code locations and the vulnerable TLS flows previously identified by \tester.

To ensure consistency and reproducibility, we leverage the same GUI agent and interaction configuration as described in Section~\ref{sec:tls_detection} when executing each app with known TMVs. This approach guarantees that the app is exercised in a manner consistent with prior vulnerability detection, maximizing the likelihood of observing relevant TLS validation events.

In the following, we detail the design of our dynamic instrumentation strategy, focusing on techniques to achieve high coverage in the presence of dynamic class loading and code obfuscation. We also describe our methodology for correlating the collected TLS validation logs with observed vulnerable flows, enabling accurate attribution of vulnerabilities to their root causes within the application's codebase.

\subject{Challenges and Strategies for Locating and Hooking TLS Validation Code via Dynamic Instrumentation.}
Our initial approach involved attempting to hook relevant classes and methods after application startup by iterating through all loaded classes and matching them against predefined rules for critical TLS validation interfaces, such as \texttt{X509TrustManager}, \texttt{WebViewClient}, and \texttt{HostnameVerifier}. However, this method suffered from low coverage due to the widespread use of dynamic class loading in Android applications, where many pertinent classes are not present at startup and are loaded only at runtime.

To address this limitation, we next experimented with hooking common API entry points. For example, to identify implementations of \texttt{X509TrustManager}, we hooked the \texttt{init} method of \texttt{SSLContext} to inspect the \texttt{TrustManager} instances provided during initialization. For \texttt{WebViewClient}, we hooked its constructor to capture all inheriting classes as they were instantiated. While these strategies improved detection rates in certain scenarios, they remained insufficient for interfaces like \texttt{HostnameVerifier}, which are implemented in a highly diverse and often application-specific manner, making them difficult to discover through such targeted entry points alone.

To address the limitations of previous approaches, we adopted the ART Tooling Interface (ART TI), a native programming interface that provides deep visibility and control over applications running on the Android Runtime (ART). ART TI enables direct access to ART's internal state, supporting advanced tasks such as profiling, debugging, and—crucially for our purposes—monitoring class loading events in real time.

By registering a callback for class load events via ART TI, our system inspects each class as it is loaded by ART. Within this callback, we evaluate whether the loaded class matches our predefined criteria, typically by checking if it implements targeted interfaces such as \texttt{HostnameVerifier} or \texttt{X509TrustManager}. When a match is found, we dynamically hook the relevant methods (e.g., \texttt{verify} for \texttt{HostnameVerifier}, \texttt{checkServerTrusted} for \texttt{X509TrustManager}). These hooks capture detailed contextual information at runtime, including the certificate being validated, the associated hostname or even full URL, and the outcome of the validation process. This rich logging enables precise correlation with the vulnerable TLS flows previously identified, allowing us to attribute validation failures to specific code segments with high confidence.

A notable constraint of ART TI is that, for security reasons, its use is generally limited to debuggable applications. To extend ART TI's capabilities to non-debuggable, production applications, we leverage Frida to hook specific system functions. This technique allows us to forcibly load the ART TI plugin by invoking its \texttt{ArtPlugin\_Initialize} method, thereby obtaining the TI environment and registering class load event callback even in restrictive environments. %

\subsubject{Summary of Instrumentation Strategies.}
In summary, our vulnerable code locator supports two distinct instrumentation strategies for locating TLS validation code: (1) leveraging the ART Tooling Interface (ART TI) to monitor class loading and dynamically hook relevant methods as classes are loaded at runtime; (2) hooking classes available at application startup and common API entry points to capture standard implementations of TLS validation interfaces. %

\subject{Precise Attribution of Vulnerable TLS Validation Code via Contextual Correlation.} 
To accurately attribute TLS validation failures to specific code segments, we record detailed execution context during each hooked method invocation. This contextual information is then correlated with the set of vulnerable TLS flows previously identified by \tester, enabling precise mapping between observed validation failures and the responsible code and library.

Our correlation strategy adapts to the nature of the available contextual data. In cases where the execution context provides explicit hostname or URL information—such as in \texttt{HostnameVerifier} callbacks (which receive the hostname as a parameter) or \texttt{WebViewClient}'s \texttt{onReceivedSslError} handler (which often has access to the relevant URL)—we directly match the Fully Qualified Domain Name (FQDN) from the runtime context to the FQDNs observed in vulnerable flows.

For contexts where such explicit domain information is unavailable, as is typical in \texttt{X509TrustManager} methods (e.g., \texttt{checkServerTrusted}), we extract the Common Name (CN) and Subject Alternative Names (SANs) from the server certificate presented during the TLS handshake. We then perform domain matching by comparing these extracted names—accounting for wildcard patterns (e.g., \texttt{*.example.com})—against the domains associated with vulnerable flows. This approach ensures robust and comprehensive correlation, even in the absence of direct hostname parameters.

While certificate-based correlation is effective, it introduces a key ambiguity that can lead to false matching. For example, suppose an application uses an \texttt{insecure} method (\textit{Code A}) for TLS flows to \textit{a.example.com} and a secure method (\textit{Code B}) for \texttt{b.example.com}. If both domains are covered by a single wildcard certificate (e.g., \textit{*.example.com}), certificate-based matching alone would incorrectly attribute the vulnerability to both \textit{Code A} and the secure \textit{Code B} for \textit{a.example.com}.

To resolve this, our code locator supports the incorporation of MitM testing during instrumentation. By actively triggering TLS validation failures and correlating them with the executing code, we can precisely identify which code path is responsible for accepting invalid certificates. In the example above, MitM testing reveals that only \texttt{Code A} accepts the invalid certificate, while \texttt{Code B} correctly rejects it. 
Similarly, it can differentiate between distinct validation paths for the same FQDN, accurately pinpointing the vulnerable code path. This active correlation eliminates ambiguity and ensures high-fidelity vulnerable code localization.

\subject{Evaluation and Results.}
To systematically evaluate the effectiveness of our vulnerable code locator, we explored four distinct configurations, varying in instrumentation strategies and the inclusion of MitM testing. Specifically, we considered two instrumentation strategies: (1) leveraging the ART Tooling Interface (ART TI) and (2) hooking common API entry points and classes available at app startup (API). For each strategy, we further assessed the impact of enabling or disabling Man-in-the-Middle (MitM) testing, resulting in a total of four configurations. Notably, we excluded the combined instrumentation approach of employing both ART TI and API hooking, as ART TI comprehensively instruments all locations that would otherwise be identified through API hooking. 

Our evaluation was conducted on 100 vulnerable applications identified by \tester. For each configuration, we measured the following metrics for flows detected by \tester: (1) \texttt{FQDN Coverage}, defined as the proportion of unique vulnerable Fully Qualified Domain Names (FQDNs) for which the responsible code could be located; (2) \texttt{Flow Coverage}, defined as the fraction of vulnerable TLS flows for which vulnerable code is located; (3) \texttt{App All}, defined as the proportion of vulnerable apps for which all detected vulnerable flows have their responsible code located; and (4) \texttt{App One}, defined as the fraction of vulnerable apps for which at least one responsible code location was found.

\begin{table}    
    \centering    
    \caption{Coverage of vulnerable code locator under different instrumentation settings.}   
    \label{tab:locator_coverage}    
    \begin{threeparttable}    
    \begin{tabular*}{\linewidth}{l@{\extracolsep{\fill}}rrrr}
        \toprule    
        \textbf{Config} & \textbf{FQDN} & \textbf{Flow} & \textbf{App All} & \textbf{App One} \\    
        \midrule    
        TI           & \textbf{36.47} & \textbf{71.88} & \textbf{14.00} & \textbf{49.00} \\    
        TI + MitM    & 30.25 & 10.27 & 9.00  & 43.00 \\    
        API          & 20.84 & 10.77 & 6.00  & 36.00 \\    
        API + MitM   & 23.44 & 9.20  & 6.00  & 30.00 \\    
        \bottomrule    
    \end{tabular*}    
    \begin{tablenotes}[flushleft]    
        \footnotesize    
            \item \textbf{Config:} TI = ART Tooling Interface; API = API/Startup Hooks; MitM = Man-in-the-Middle testing enabled.
    \end{tablenotes}    
    \end{threeparttable}    
\end{table}

\subsubject{Observations.}
As shown in Table~\ref{tab:locator_coverage}, the ART TI-based instrumentation consistently outperforms the API hooking approach across all coverage metrics. In the non-MitM configuration, the TI strategy achieves significantly higher \texttt{FQDN Coverage} (36.47\% vs. 20.84\%) and a substantially greater \texttt{Flow Coverage} (71.88\% vs. 10.77\%) compared to the API strategy. While enabling MitM tends to reduce the coverage of previously identified flows—most notably dropping the \texttt{Flow Coverage} for TI from 71.88\% to 10.27\%—the TI-based approach maintains its lead over API hooking in the MitM configuration as well.

\subject{Analysis of Localization Failures.} To understand why our tool was unable to locate vulnerable code for certain apps and FQDNs, we conducted targeted reverse engineering and manual testing. Our investigation identified several root causes for these localization failures: (1) Untriggered Execution Paths: Certain vulnerable code segments were not executed during dynamic analysis. This occurred either because the automated interaction script did not cover the specific user interface actions required to activate the functionality, or because applications employed anti-analysis mechanisms that led to premature termination upon detecting the instrumentation environment. (2) Native Code Implementations: A significant portion of failures involved networking logic implemented in native code. Locating the source of vulnerabilities becomes particularly challenging when these native libraries are stripped of their debugging symbols, as seen in some custom or stripped builds of networking libraries. (3) Non-Standard Implementations: Some applications utilize unconventional or proprietary protocols and libraries for handling TLS, which fall outside the scope of the standard APIs and libraries targeted by our instrumentation.

\subject{Deployment Setup.} 
Based on the evaluation results, we adopt the ART Tooling Interface (ART TI) as the instrumentation strategy for its superior coverage and precision. For deployment, each vulnerable app is executed with Man-in-the-Middle (MitM) testing enabled. As discussed previously, this active correlation is crucial for ensuring high-fidelity localization by precisely pinpointing the insecure code path responsible for a validation failure, thus resolving ambiguities that arise when multiple code paths handle traffic to the same destination. This process directly locates the vulnerable TLS code snippets that are accurately correlated with observed failures.

\subsection{Vulnerable Code Classifier of \analyzer}
\label{subsec:rca_code_classify}

Given the 8,374 vulnerable apps identified by \tester, applying our vulnerable code locator uncovered 8,065 distinct instances of vulnerable code—typically, functions responsible for TLS certificate validation. This corresponds to a coverage of 51.28\% of vulnerable apps and 33.81\% of app-specific TLS FQDNs. As a key step in our root cause analysis, we next perform automated, fine-grained categorization of these vulnerable code segments, which constitutes the vulnerable code classification module of \analyzer.

\subject{Comprehensive Taxonomy of TLS Validation Code Errors.} As discussed in Section~\ref{sec:pre}, prior research has rarely explored code-level categories of TLS validation errors, and existing studies are limited to manual case studies with coarse-grained categorizations. To advance the state of the art, we develop a comprehensive and fine-grained taxonomy of common error patterns in TLS validation code, enabling systematic root cause analysis.

Our taxonomy was constructed through an iterative process that combine automated analysis using Large Language Models (LLMs) and expert manual review. We began by applying DeepSeek-V3 to a representative subset of vulnerable code snippets, using a basic prompt to assign them to broad and corse-grained categories—such as empty or insecure \texttt{TrustManager}, insecure \texttt{WebViewClient}, and insecure \texttt{HostnameVerifier}. These initial results were then manually validated and refined, with additional rounds of LLM-based classification and human review to capture more nuanced and recurring error patterns. This process resulted in a detailed set of error categories tailored to the main interfaces responsible for TLS validation in Android: \texttt{X509TrustManager}, \texttt{WebViewClient}, and \texttt{HostnameVerifier}. 

Table~\ref{tab:category_definition} summarizes our final taxonomy, which captures both broad and nuanced distinctions in insecure TLS validation logic. For \texttt{X509TrustManager}, categories span from completely empty implementations (T1) that bypass all certificate checks, to non-empty but insecure variants (T2) that perform only superficial or incorrect validation—such as checking only certificate validity periods (T2-A), verifying parameter presence (T2-B), or inspecting subject fields without proper chain validation (T2-C). Additional subcategories address flawed signature checks (T2-D), exception suppression (T2-E), and conditional or incomplete logic (T2-F). The taxonomy also includes a secure baseline (T0) and an “unknown” category (TU) for unclassifiable or obfuscated code. 

For \texttt{WebViewClient}, the taxonomy focuses on how SSL errors are handled in \texttt{onReceivedSslError}, distinguishing between unconditional acceptance of invalid certificates (W1), conditional acceptance based on user prompts or specific error types (W2 and its subcategories), secure handling (W0), and an “unknown” category (WU) for complex or obfuscated logic. \texttt{HostnameVerifier} categories differentiate secure implementations (H0) and those that always return true (H1). Insecure but not-always-true implementations (H2) are further classified based on their flawed logic, such as incorrect hostname-based validation (H2-A) or incomplete subject matching (H2-B). An “unknown” category (HU) is reserved for unclassifiable implementations.

Each category reflects a distinct root cause, enabling precise attribution of vulnerabilities and supporting targeted remediation. For example, T1 and H1 represent cases where security checks are entirely disabled, while T2 and its subcategories capture a range of partial or misguided validation attempts. Conditional categories in \texttt{WebViewClient} (W2) highlight the risks of delegating security decisions to users or relying on application state, which can result in inconsistent or unsafe behavior.

\begin{table*}
\centering
\caption{Performance of different in-context learning (ICL) settings for TLS error code categorization.}
\label{tab:llm_categorization_performance}
\begin{threeparttable}
\begin{tabular}{lcccc}
\toprule
\textbf{Setting} & \textbf{All Categories} & \textbf{T2 Subcategories} & \textbf{W2 Subcategories} & \textbf{H2 Subcategories} \\
\midrule
P1 (DeepSeek, zero-shot) & 0.76 / 0.82 / 0.79 & 0.70 / 0.64 / 0.65 & 1.00 / 0.90 / 0.95 & 0.44 / 0.24 / 0.31 \\
P1 (DeepSeek, 6-shot) & 0.94 / 0.96 / 0.95 & 0.78 / 0.86 / 0.81 & 0.96 / 0.95 / 0.95 & 1.00 / 0.88 / 0.94 \\
P1 (Qwen2.5-72B, 6-shot) & 0.94 / 0.97 / 0.95 & 0.75 / 0.91 / 0.81 & 0.93 / 0.90 / 0.91 & 1.00 / 0.94 / 0.97 \\
P2 (DeepSeek, zero-shot) & 0.80 / 0.83 / 0.80 & 0.72 / 0.64 / 0.66 & 1.00 / 0.90 / 0.95 & 0.87 / 0.47 / 0.60 \\
P2 (DeepSeek, 6-shot) & 0.94 / 0.95 / 0.94 & 0.74 / 0.83 / 0.77 & 0.96 / 0.95 / 0.95 & 0.93 / 0.82 / 0.87 \\
P2 (DeepSeek, 10-shot) & 0.96 / 0.96 / 0.96 & 0.86 / 0.89 / 0.87 & 0.96 / 0.95 / 0.95 & 0.93 / 0.82 / 0.87 \\
P2 (DeepSeek, 10-shot, with comment) & 0.97 / 0.97 / 0.97 & 0.92 / 0.90 / 0.90 & 0.96 / 0.95 / 0.95 & 1.00 / 0.88 / 0.94 \\
\bottomrule
\end{tabular}
\begin{tablenotes}[flushleft]
    \footnotesize
    \item[*] Scores are reported as Precision / Recall / F1-score.
    \item [*] The main differences between prompt versions P1 and P2 are that P2 explicitly includes "Unknown" or umbrella categories, allowing the model to handle unclassifiable cases.
\end{tablenotes}
\end{threeparttable}
\end{table*}

To support rigorous evaluation of our LLM-based classification approach, we also curated a benchmark dataset comprising 365 distinct Java classes, each exhibiting one or more of the identified TLS validation vulnerabilities. This dataset serves as a representative and diverse testbed for assessing the accuracy and robustness of automated categorization methods.

\begin{table*}
    \caption{Categories for TLS Validation Code Errors}
    \label{tab:category_definition}
    \centering
    \begin{tabular}{l p{0.8\textwidth}}
        \toprule
        \multicolumn{2}{l}{\textbf{For \texttt{X509TrustManager} implementations:}} \\[1ex]
        \textbf{T0} & \textbf{Secure TrustManager.} The implementation correctly performs all necessary certificate validation checks according to standard security practices. \\[0.5ex]
        \textbf{T1} & \textbf{Empty TrustManager.} The core validation methods (e.g., \texttt{checkServerTrusted}, \texttt{checkClientTrusted}) are implemented with empty bodies or contain only minimal logic that effectively bypasses all certificate checks. \\[0.5ex]
        \textbf{T2} & \textbf{Non-Empty but Insecure TrustManager.} The implementation attempts some validation, but the logic is flawed or incomplete, rendering it insecure. \\
        & This category includes the following generally mutually exclusive sub-types: \\
        \textbf{T2-A} & \textbf{Only Checked Validity Period.} The validation logic is limited to verifying the certificate's \texttt{notBefore} and \texttt{notAfter} dates, neglecting other critical checks such as signature integrity. \\
        \textbf{T2-B} & \textbf{Only Checked if Parameters were Empty or Null.} The implementation primarily checks for the presence or non-nullity of input parameters (e.g., the certificate chain or authentication type) without performing actual cryptographic validation of the certificates themselves. \\
        \textbf{T2-C} & \textbf{Only Checked the Certificate's Subject.} The validation logic is confined to inspecting fields within the certificate's subject, such as common name or organization, without proper verification of the certificate chain or its signature. \\
        \textbf{T2-D} & \textbf{Verified Signature but Not Certificate Chain.} The implementation may check the signature of the leaf certificate or an intermediate certificate but fails to validate the integrity and trustworthiness of the entire certificate chain up to a trusted root anchor. \\
        & Additionally, two non-mutually exclusive sub-types can apply to T2, potentially co-occurring with sub-types T2-A through T2-D, indicating broader patterns of insecurity: \\
        \textbf{T2-E} & \textbf{Ignored Certificate Validation Exception.} The code explicitly catches certificate validation-related exceptions (e.g., \texttt{CertificateException}, \texttt{NoSuchAlgorithmException}) and proceeds as if validation succeeded, often by having an empty catch block, merely logging the error, or returning without signaling failure. \\
        \textbf{T2-F} & \textbf{Verified Certificates Only Under Limited Conditions.} The validation logic is conditional and flawed, for example, by trusting specific, potentially insecure, issuers or certificates based on hardcoded values, or by applying correct checks only in very narrow, often insecurely defined, circumstances or for specific domains. \\[0.5ex]
        \textbf{TU} & \textbf{Unknown TrustManager.} The code is too complex, heavily obfuscated, or employs unconventional or proprietary validation mechanisms that prevent its classification into any of the predefined categories. \\[1ex]
        
        \midrule
        \multicolumn{2}{l}{\textbf{For \texttt{WebViewClient} implementations:}} \\[1ex]
        \textbf{W0} & \textbf{Secure WebViewClient.} The \texttt{onReceivedSslError} method handles SSL errors appropriately, typically by calling \texttt{handler.cancel()} to terminate the connection, or by presenting an informative error to the user without an option to proceed insecurely. \\[0.5ex]
        \textbf{W1} & \textbf{Completely Ignore SSL Errors.} The implementation unconditionally calls \texttt{handler.proceed()}, instructing the WebView to ignore the SSL error and load the content from the insecure connection without any user intervention or warning. \\[0.5ex]
        \textbf{W2} & \textbf{Conditional Ignore SSL Error.} The decision to ignore SSL errors by calling \texttt{handler.proceed()} depends on certain conditions: \\
        \textbf{W2-A} & \textbf{Pop-up Warning with Option to Ignore Error.} The application presents a dialog to the user, warning about the SSL error but providing an explicit option for the user to proceed despite the risk. \\
        \textbf{W2-B} & \textbf{Ignored Specific Error Types.} The implementation calls \texttt{handler.proceed()} only for certain types of SSL errors (e.g., based on SslError codes), potentially misinterpreting the severity or ignoring critical errors that should terminate the connection. \\
        \textbf{W2-C} & \textbf{Ignored Error When the App is in a Specific State.} SSL errors are ignored if the application is in a particular state, such as a ``debug mode,'' a globally configured ``insecure connection allowed'' setting, or when connecting to specific (potentially developer-controlled or internal) domains where SSL errors are anticipated or intentionally bypassed. \\[0.5ex]
        \textbf{WU} & \textbf{Unknown WebViewClient.} The logic within \texttt{onReceivedSslError} is unclassifiable using the defined categories due to its complexity, obfuscation, or unusual error handling patterns. \\[1ex]
        
        \midrule
        \multicolumn{2}{l}{\textbf{For \texttt{HostnameVerifier} implementations:}} \\[1ex]
        \textbf{H0} & \textbf{Secure HostnameVerifier.} The \texttt{verify} method correctly implements hostname verification according to standard practices, ensuring the hostname in the URL matches the hostname(s) in the server's certificate. \\[0.5ex]
        \textbf{H1} & \textbf{Always True Hostname Verification.} The \texttt{verify} method unconditionally returns \texttt{true}, effectively disabling hostname verification and allowing any certificate to be accepted for any hostname. \\[0.5ex]
        \textbf{H2} & \textbf{Not Always True but Insecure Verification.} The \texttt{verify} method does not always return \texttt{true} but contains flawed logic that may accept incorrect hostnames under certain conditions. This category includes the following sub-types: \\
        \textbf{H2-A} & \textbf{Incorrect Hostname-Based Validation.} The validation logic is based on checks against the hostname parameter itself instead of checking the hostnames provided in the certificate's subject. \\
        \textbf{H2-B} & \textbf{Incomplete Subject Matching.} The code attempts to match the hostname against the certificate's subject, but the logic is flawed, for example, by performing partial or incorrect matches, or by improperly parsing the subject. \\
        \textbf{HU} & \textbf{Unknown HostnameVerifier.} The hostname verification logic is unclassifiable due to its complexity, obfuscation, or reliance on non-standard mechanisms. \\
        \bottomrule
    \end{tabular}
\end{table*}

\subject{LLM-based Classification of TLS Validation Code Errors.} 
Building on our comprehensive taxonomy, we employ Large Language Models (LLMs) to automatically classify the extracted vulnerable code snippets. For each snippet, the LLM receives both the code and the relevant set of category definitions based on the implemented interface (e.g., \texttt{X509TrustManager} categories for TrustManager code). The LLM is then tasked with assigning the most appropriate category label(s) to the snippet.

To ensure robust and accurate classification, we systematically evaluated a range of LLM-based in-context learning strategies. Our experiments compared different foundation models, prompt enhancement techniques—including the addition of explanatory comments to in-context examples—and the prompt templates detailed in Appendix~\ref{appendix:classifier_prompts}. We also varied the number of in-context examples to assess their impact on classification performance. This iterative process allowed us to identify the optimal configurations for high-precision, scalable categorization of TLS validation errors.

 To rigorously assess the effectiveness of our LLM-based classification approach, we utilized the benchmark dataset described above. To further increase the diversity and representativeness of the evaluation set, we manually rewrote a subset of code snippets, ensuring that their underlying semantic vulnerabilities were preserved. Table~\ref{tab:llm_categorization_performance} presents the classification results across various in-context learning (ICL) configurations, reporting Precision, Recall, and F1-score for all categories, as well as for the more challenging T2 (\texttt{TrustManager}), W2 (\texttt{WebViewClient}), and H2 (\texttt{HostnameVerifier}) subcategories. This comprehensive evaluation demonstrates the robustness and accuracy of our automated categorization pipeline under different prompt and model settings.

Several key observations emerge from the results in Table~\ref{tab:llm_categorization_performance}. First, increasing the number of few-shot examples enhances classification performance. The addition of explanatory comments to in-context learning (ICL) examples, as demonstrated in the "P2 (DeepSeek, 10-shot, with comment)" setting, further improves performance—particularly for the challenging T2 subcategories, where the F1-score rises from 0.87 to 0.90. Notably, the introduction of umbrella categories (TU, WU, HU) in the P2 prompt does not significantly impact performance on well-defined categories, while providing essential flexibility for ambiguous or complex cases. 

Based on these findings, our final LLM-based categorization pipeline uses the DeepSeek-V3 model with a few-shot learning strategy. It employs 10 examples, some of which include explanatory comments to clarify the rationale for their ground-truth labels.

\subsection{Analysis of Vulnerable Code Snippets}
\label{subsec:rca_result}
\begin{figure*}
    \centering
    \includegraphics[width=.98\textwidth]{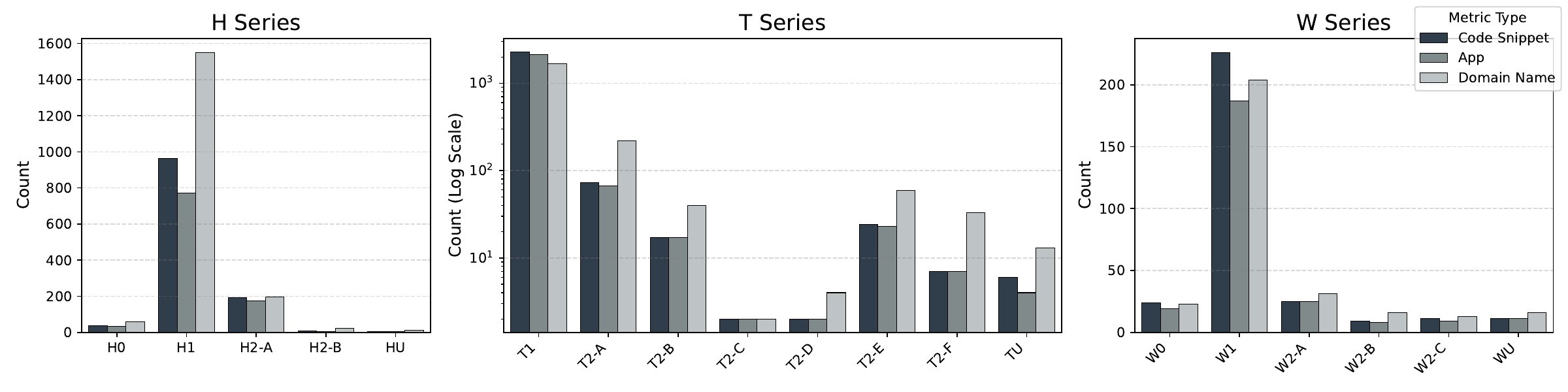}
    \caption{Distribution of vulnerable TLS validation code snippets across fine-grained categories defined in Table~\ref{tab:category_definition}. What is also presented are the category distribution of Android apps (App) and TLS FQDNs (Domain Name). 
    }
    \label{fig:category_distribution_vul_dedup}
\end{figure*}

Leveraging the LLM-based classifier described in Section~\ref{subsec:rca_code_classify}, we systematically categorized the vulnerable TLS validation code snippets. In total, the locator extracted 7,375 code snippets for 3,944 apps and 3,565 TLS FQDNs. Since the same library may be present in multiple apps, we performed deduplication based on fully qualified class names, resulting in 3,904 unique code snippets. It is important to note that, due to code obfuscation and variations in class naming, complete deduplication is not always possible.

\subject{Category Distribution.} Figure~\ref{fig:category_distribution_vul_dedup} presents distribution of these vulnerable code snippets across defined categories. The figure also shows the category distribution at both the app and TLS FQDN levels. For apps, an app is counted under a category if at least one of its vulnerable code snippets falls into that category. Similarly, for TLS FQDNs, an FQDN is counted for a category if its corresponding code snippet is classified as such. This comprehensive categorization provides a clear view of the most prevalent types of TLS validation errors in real-world Android applications, highlighting both the diversity and concentration of insecure coding practices across the ecosystem.

Our analysis reveals several prevalent patterns of TLS validation errors. Among \texttt{X509TrustManager} implementations, the T1 category, where methods like \texttt{checkServerTrusted} simply return without any checks, is overwhelmingly common. This pattern is observed in a substantial number of unique classes, affecting a large cohort of applications and numerous distinct TLS domain names, indicating a widespread practice of completely bypassing server certificate validation. It also motivates our study to investigate why the responsible party (the developers) chose this kinds of apparenty insecure implementation, which will be detailed next in Subsection~\ref{subsec:rca_party}.

The T2 category encompasses flawed but not entirely empty implementations. For instance, T2-A represents code that solely invokes \texttt{certificate.checkValidity()}, neglecting crucial chain and issuer verification. This is a frequently observed sub-pattern within T2. Another notable sub-category, T2-E, involves suppressing \texttt{CertificateException} or other critical exceptions within a \texttt{try-catch} block, effectively neutralizing any underlying validation attempt. Other T2 sub-categories, though less frequent, highlight specific misunderstandings: T2-B implementations only perform rudimentary checks for null or empty parameters; T2-C incorrectly relies on string matching within the certificate's subject; and T2-D might attempt to verify a certificate's signature but fails to validate the entire certificate chain up to a trusted root. Conditional bypasses, categorized as T2-F, where validation is skipped under certain conditions (e.g., based on specific auth types), also contribute to the landscape of insecure trust managers. Figure~\ref{fig:category_code_examples} presents three code examples falling into the T categories. Among them, T2-A only checks the certificate's validity period, T2-B only checks if parameters are empty or null, and T2-D incorrectly verifies a certificate’s signature without validating the chain.

\begin{figure*}[htb]
    \newsavebox{\mycodebox}
    \newenvironment{framedcode}[1][4cm]{%
        \begin{lrbox}{\mycodebox}%
        \begin{minipage}[t][#1][t]{\dimexpr\linewidth-2\fboxsep-2\fboxrule\relax}%
    }{%
        \end{minipage}%
        \end{lrbox}%
        \fbox{\usebox{\mycodebox}}%
    }

    \lstset{
        language=Java,
        breaklines=true,
        basicstyle=\ttfamily\scriptsize,
        aboveskip=0pt,
    }
    \captionsetup[subfigure]{skip=3pt}
    \centering  

    \begin{subfigure}[t]{0.31\textwidth}  
        \centering
        \begin{framedcode}
        \begin{lstlisting}
public void checkServerTrusted(X509Certificate[] chain, String authType)
{
    chain[0].checkValidity();
    return;
}
        \end{lstlisting}
        \end{framedcode}
        \caption{T2-A: Only checked validity period}  
    \end{subfigure}  
    \hfill  
    \begin{subfigure}[t]{0.31\textwidth}  
        \centering
        \begin{framedcode}
        \begin{lstlisting}
public void checkServerTrusted(X509Certificate[] chain, String authType)
{
  if (chain == null || chain.length <= 0) {
    throw new IllegalArgumentException();
  }
}
        \end{lstlisting}
        \end{framedcode}
        \caption{T2-B: Only checked if the parameters were empty or null}  
    \end{subfigure}  
    \hfill  
    \begin{subfigure}[t]{0.31\textwidth}  
        \centering
        \begin{framedcode}
        \begin{lstlisting}
public void checkServerTrusted(X509Certificate[] chain, String authType)
{
    int i = 0;
    while (i < chain.length) {
        X509Certificate cert = chain[i];
        cert.verify(cert.getPublicKey());
        i++;
    }
    return;
}
        \end{lstlisting}
        \end{framedcode}
        \caption{T2-D: Verified signature but not certificate chain}  
    \end{subfigure}  
    \vspace{1.5ex}
    
    \begin{subfigure}[t]{0.31\textwidth}  
        \centering
        \begin{framedcode}
        \begin{lstlisting}
public boolean verify(String hostname, SSLSession session)
{
    if ("example.com".equals(hostname)) {
        return 1;
    } else {
        return 0;
    }
}
        \end{lstlisting}
        \end{framedcode}
        \caption{H2-A: Incorrect use of hostname for validation}  
    \end{subfigure}  
    \hfill  
    \begin{subfigure}[t]{0.31\textwidth}  
        \centering
        \begin{framedcode}
        \begin{lstlisting}
public void onReceivedSslError(WebView view, SslErrorHandler handler, SslError error)
{
    error = error.getPrimaryError();
    if ((error == 3) || (error == 5)) {
        handler.proceed();
    } else {
        handler.cancel();
    }
    return;
}
        \end{lstlisting}
        \end{framedcode}
        \caption{W2-B: Ignored specific error types}  
    \end{subfigure}  
    \hfill  
    \begin{subfigure}[t]{0.31\textwidth}  
        \centering
        \begin{framedcode}
        \begin{lstlisting}
public void onReceivedSslError(WebView view, SslErrorHandler handler, SslError error)
{
    if (!Config.isCheckTrustCert()) {
        handler.proceed();
    } else {
        handler.cancel();
    }
    return;
}
        \end{lstlisting}
        \end{framedcode}
        \caption{W2-C: Ignored error when the app is in a specific state}  
    \end{subfigure}  
    \caption{Representative code examples for three major TLS validation error categories: (T) TrustManager, (H) HostnameVerifier, and (W) WebViewClient. Each pair illustrates a distinct error pattern as defined in Table~\ref{tab:category_definition}.}  
    \label{fig:category_code_examples}  
\end{figure*}

Regarding \texttt{WebViewClient} implementations, overriding \texttt{onReceivedSslError} to insecurely handle SSL errors is a common vulnerability. The W1 category, where \texttt{handler.proceed()} is called unconditionally, is particularly widespread, impacting a significant number of applications and a broad range of domains. This indicates a prevalent tendency to prioritize uninterrupted browsing experience over security. The W2 category also presents considerable risk. Sub-category W2-A, where users are prompted but can choose to proceed despite an SSL error (e.g., displaying an alert dialog with "Continue" and "Cancel" options), shifts the security decision to potentially uninformed users. Other conditional bypasses include W2-B, ignoring specific types of SSL errors, and W2-C, ignoring errors based on application state, such as a debug flag. Figure~\ref{fig:category_code_examples} presents two code examples falling into the W categories. Among them, W2-B demonstrates ignoring specific error types, while W2-C shows ignoring an error based on the application's configuration state.

For \texttt{HostnameVerifier} implementations, the H1 category, characterized by a \texttt{verify} method that unconditionally returns \texttt{true}, is a clear and frequently identified vulnerability. This effectively disables the check that ensures the connected hostname matches the one specified in the server's certificate. While less frequent than H1, the H2-A category is also a significant issue, involving incorrect hostname-based validation where checks are performed against the hostname parameter itself, instead of comparing it with the hostnames in the certificate's Common Name and Subject Alternative Names. Less common, but still problematic, are H2-B implementations, which attempt some form of hostname matching but do so using flawed logic, such as overly permissive wildcard matching or insecure substring checks. Figure~\ref{fig:category_code_examples} presents a vulnerable code example falling into the H categories. The H2-A example shows a flawed implementation that validates the hostname parameter rather than the certificate subject.

It is noteworthy that secure categories (W0 and H0) also contribute some counts. These occurrences primarily stem from two sources. First is methodological, stemming from our domain-based approach for matching vulnerable network flows to the instrumented code. For instance, if a specific domain, identified as vulnerable due to one insecure validation path, is also processed by a genuinely secure validation routine within the same application, our matching logic might associate the secure code snippet with the vulnerable flow observed for that domain.

Second, wrapper implementations that delegate validation to another verifier can also contribute to these counts. For instance, a \texttt{HostnameVerifier} implementation might simply call the system's default verifier, e.g., via \texttt{getDefaultHostnameVerifier()}. While the wrapper itself correctly delegates the check and is thus classified as secure (H0), a vulnerability can still arise if the default verifier has been insecurely modified elsewhere by the app or a third-party SDK. Although this delegation pattern is theoretically possible for \texttt{X509TrustManager} implementations (T0) as well, we found it to be rare in practice, with no such cases identified in our dataset. On the other hand, we located several instances of insecure H0 arising from this delegation pattern. Figure~\ref{fig:insecure-h0-example} shows an example where a class implements a seemingly secure \texttt{HostnameVerifier} by calling the system's default verifier, but this default is overridden elsewhere in the application with an insecure, always-true implementation (H1).

\begin{figure}[t]
\centering
\begin{lstlisting}[language=Java, caption={A seemingly secure \texttt{HostnameVerifier} that delegates to the default.}, label={lst:delegating-verifier}, basicstyle=\ttfamily\scriptsize, breaklines=true, showstringspaces=false]
public class DelegatingVerifier implements HostnameVerifier {
    @Override
    public boolean verify(String hostname, SSLSession session) {
        // This implementation appears secure as it delegates the
        // check to the system's default verifier.
        HostnameVerifier defaultVerifier = 
            HttpsURLConnection.getDefaultHostnameVerifier();
        return defaultVerifier.verify(hostname, session);
    }
}
\end{lstlisting}
\vspace{0.5em}
\begin{lstlisting}[language=Java, caption={Insecure setup code elsewhere in the app that compromises the default verifier.}, label={lst:insecure-setup}, basicstyle=\ttfamily\scriptsize, breaklines=true, showstringspaces=false]
public class MainApplication extends android.app.Application {
    @Override
    public void onCreate() {
        super.onCreate();
        
        // The default verifier is overridden with an insecure,
        // always-true (H1) implementation, affecting all callers.
        HttpsURLConnection.setDefaultHostnameVerifier(
            new HostnameVerifier() {
                @Override
                public boolean verify(String hostname, SSLSession session) {
                    return true; // H1: Always returns true
                }
            }
        );
    }
}
\end{lstlisting}
\caption{An example of an insecure H0 classification. The \texttt{DelegatingVerifier} in (a) appears secure by correctly delegating to the system's default verifier. However, separate application initialization code in (b) overrides this default with an insecure, always-true (H1) implementation, rendering the delegation in (a) vulnerable.}
\label{fig:insecure-h0-example}
\end{figure}

\begin{figure*}[htb]
    \centering
    \includegraphics[width=.98\textwidth]{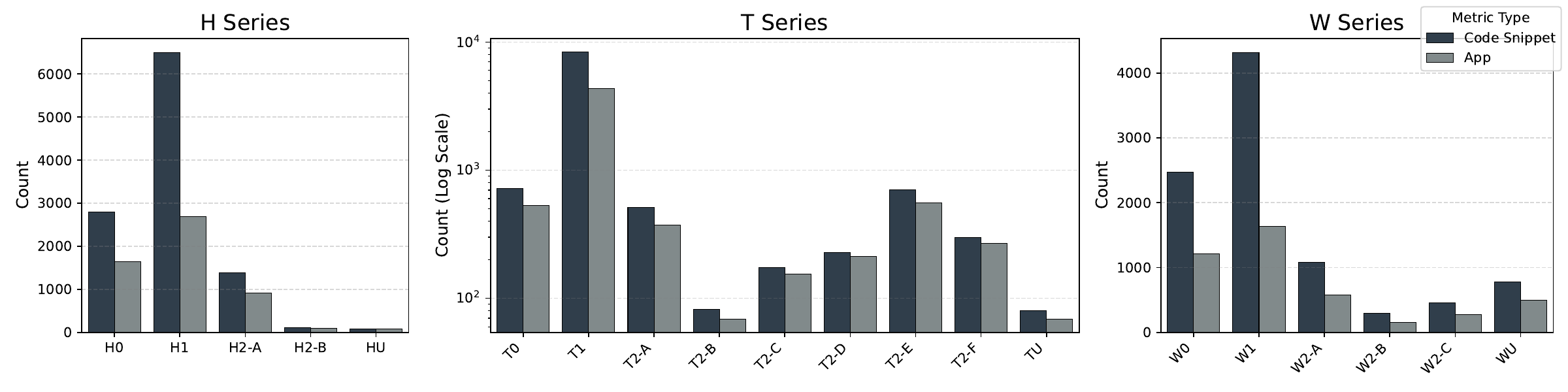}
    \caption{Distribution of TLS validation error categories for 31,177 unique classes from vulnerable APPs in our dataset, detailing affected methods and apps.}
    \label{fig:category_distribution_all_dedup}
\end{figure*}

\subject{Applying the Classifier to TLS Valdation Code Extracted via Static Analysis.}
Leveraging dynamic instrumentation, our vulnerable code locator can enable precise correlation but is at the expense of limited efficiency and coverage. To complement its inherently incomplete coverage, we also applied rule-based static analysis across the vulnerable apps. This process extracts all available TLS validation code snippets, regardless of whether their correlation with a specific vulnerable flow is known.
This process yields 31,177 distinct TLS validation code snippets. Further categorization using our LLM-based classifier reveals a distribution depicted in Figure~\ref{fig:category_distribution_all_dedup}. As we can see, the distribution largely corroborates the patterns observed from the instrumentation-based samples, reinforcing the prevalence of issues like T1 and W1.

\begin{figure*}[htb]
    \centering
    \includegraphics[width=.98\textwidth]{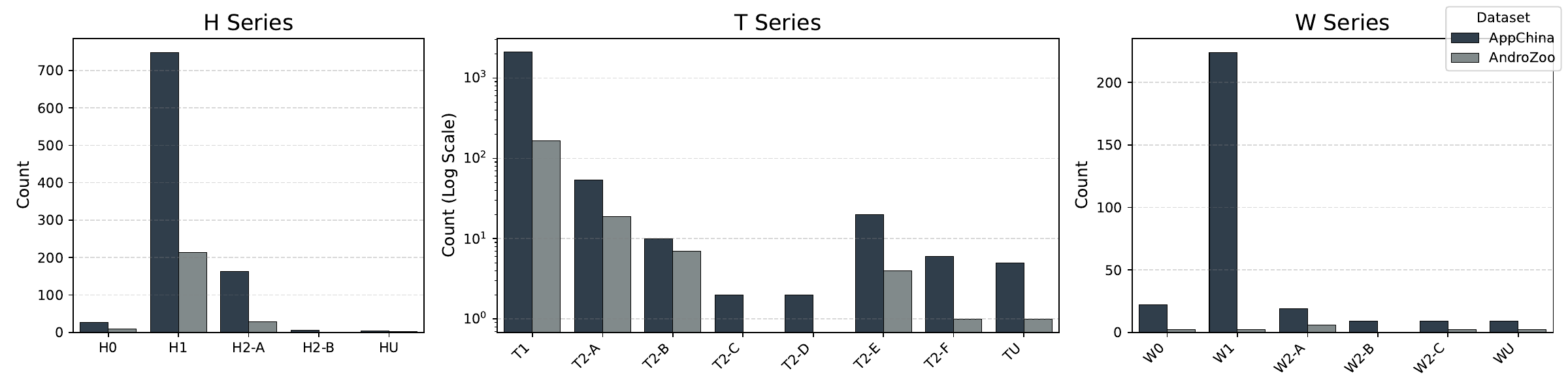}
    
    \caption{Comparison of TLS validation error category distributions between AppChina and Androzoo apps (deduplicated by class name).}
    \label{fig:category_distribution_appchina_androzoo}
\end{figure*}

\subject{Difference in Error Code Category Distribution between AppChina apps and Androzoo Apps.} We also compare the distribution of TLS validation error categories between vulnerable apps from the AppChina dataset and those from the Androzoo dataset (i.e. Google Play apps). Figure~\ref{fig:category_distribution_appchina_androzoo} illustrates the relative proportions of each error category for both datasets, based on deduplicated code snippets.

Our analysis reveals that while the overall patterns are similar—H1 (Always True Hostname Verification) and T1 (Empty TrustManager) remain the most prevalent categories in both datasets—there are several notable differences. The most striking distinction is in the W1 (Completely Ignore SSL Errors) category, which is a major vulnerability type in the AppChina dataset but is nearly absent from the AndroZoo samples. Additionally, some specific insecure TrustManager implementations, namely T2-C (Only Checked the Certificate's Subject) and T2-D (Verified Signature but Not Certificate Chain), appear in the AppChina dataset but are entirely absent from AndroZoo. The distribution patterns for most other categories are broadly comparable between the two datasets.

These differences may reflect variations in development practices, library usage, or security awareness between the two app ecosystems. Nonetheless, the dominance of the most severe error patterns across both datasets underscores the widespread and persistent nature of insecure TLS validation practices in the Android ecosystem.

\begin{figure}
    \centering
    \includegraphics[width=0.48\textwidth]{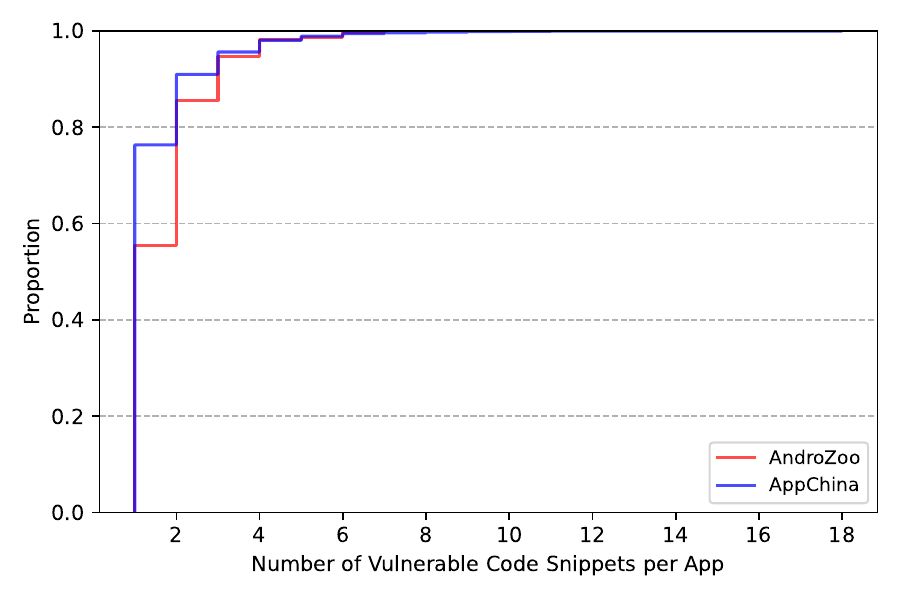}
    \caption{CDF of apps by the number of unique vulnerable TLS validation code snippets per app.}
    \label{fig:app_vul_snippet_dist}
\end{figure}

\subject{To What Extent Can an App Be Jointly Affected by Multiple Vulnerable Code Snippets for TLS Validation?} To gain deeper insights into the prevalence and concentration of TLS validation vulnerabilities within individual applications, we analyze the distribution of apps based on the number of unique vulnerable code snippets identified per app. Figure~\ref{fig:app_vul_snippet_dist} illustrates the cumulative distribution function (CDF) of apps with respect to the number of deduplicated vulnerable code snippets extracted via dynamic instrumentation.

Our results show that a significant fraction of vulnerable apps contain only a single unique vulnerable code snippet. Specifically, 54.45\% of vulnerable apps from AndroZoo and 66.33\% from AppChina feature exactly one such snippet. This indicates that in many cases, a single insecure implementation is responsible for all observed TMVs within the app. Furthermore, the cumulative distribution reveals that 83.99\% of vulnerable AndroZoo apps and 87.92\% of AppChina apps have two or fewer unique snippets. The presence of multiple distinct vulnerable code snippets in a non-trivial portion of apps suggests the use of different insecure libraries, custom implementations, or duplicated insecure logic. This highlights the importance of comprehensive code analysis, as addressing only one instance may leave other vulnerable paths unmitigated within the same application.

\subject{To What Extent Can a Single App-specific Vulnerable Code Snippet Affect Different FQDNs and Apex Domains?} To better understand the relationship between vulnerable code snippets and affected TLS FQDNs and apex domains, we analyze how vulnerable validation code snippets are distributed across these dimensions. 

We observe that a single vulnerable code snippet is frequently responsible for TMVs affecting multiple FQDNs and apex domains. In many cases, a single insecure TrustManager or HostnameVerifier implementation is used for all outbound TLS connections within an app, resulting in broad exposure across both FQDNs and their corresponding apex domains. However, there are also instances where different FQDNs or apex domains within the same app are associated with distinct vulnerable code snippets, reflecting the coexistence of multiple libraries or custom implementations.

\begin{figure}
    \centering
    \includegraphics[width=0.48\textwidth]{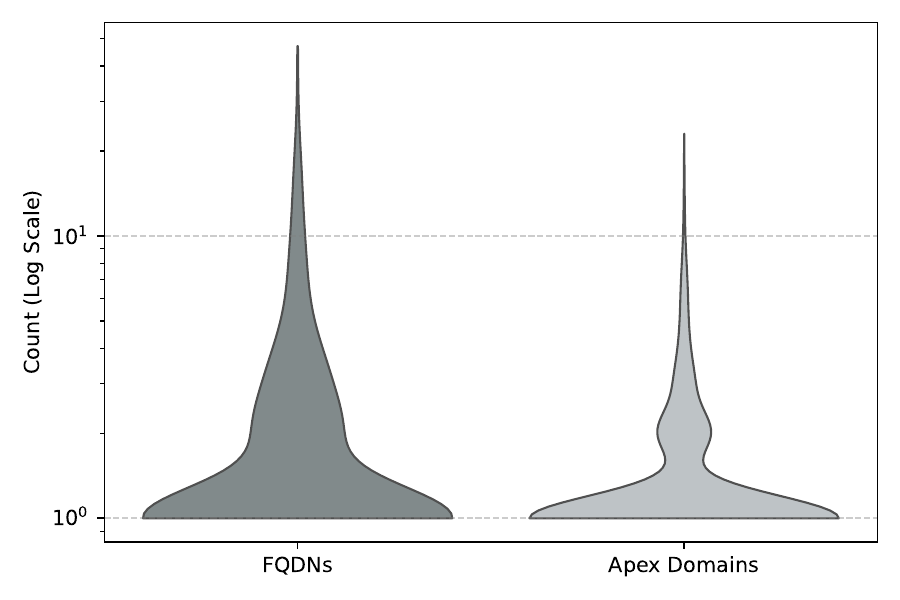}
    
    \caption{Distribution of the number of TLS FQDNs and apex domains associated with each unique vulnerable code snippet.}
    \label{fig:snippet_fqdn_app_dist}
\end{figure}

Figure~\ref{fig:snippet_fqdn_app_dist} summarizes these distributions, showing the number of FQDNs and apex domains per vulnerable code snippet. The long-tailed nature of these distributions highlights both the pervasiveness of code reuse and the diversity of insecure validation logic in the Android ecosystem.

\subsection{Responsible Party Attribution}
\label{subsec:rca_party}
To understand which parties are primarily responsible for TLS MitM vulnerabilities (TMVs) and especial the vulnerable code snippets in Android apps, we systematically attribute each vulnerable code snippet to either the app developer or a specific third-party library provider. This attribution is crucial for effective vulnerability disclosure and for prioritizing mitigation efforts.

\subject{Methodology for Responsible Party Attribution.}
We begin by extracting the package names associated with each vulnerable code snippet identified and categorized by \analyzer. To distinguish between first-party (app developer) and third-party code, we compile a list of the most frequently occurring package name prefixes across all deduplicated vulnerable snippets. We focus on package names that appear in at least two distinct apps, as these are strong indicators of third-party library usage.

Each candidate package name is manually annotated with three key attributes: (1) whether it belongs to a third-party library, (2) the specific library name (e.g., ``cn.jiguang.net'' corresponds to JPush), and (3) the provider, i.e., the organization or company responsible for maintaining the library. To streamline the annotation process, we first cross-reference package names with the Google Play SDK Index~\footnote{https://play.google.com/sdks}, which catalogs widely used SDKs and their respective providers. For package names not found in the index, we conduct manual verification by inspecting the codebase and performing targeted Internet searches to accurately determine their origin and function.

Through this systematic approach, we annotate 52 out of 235 unique package names as belonging to third-party libraries, collectively accounting for 35.27\% of all vulnerable code snippets. In total, we identify 38 distinct third-party libraries maintained by 26 unique providers. This comprehensive labeling enables precise attribution of vulnerabilities and supports targeted remediation efforts across the Android ecosystem.

Notably, there are still packages for which we were unable to confirm their respective library and provider, leaving them excluded from the third-party set. Therefore, the statistics of third-party libraries and providers reported in this study represent lower-bound estimates.

\subject{To What Extent Do Third-party Libraries Account for TMVs?}
Table~\ref{tab:party_attribution_summary} presents a detailed breakdown of vulnerable code snippets, affected apps, and impacted TLS FQDNs by responsible party. Our analysis reveals that approximately 41\% of all vulnerable code snippets originate from third-party libraries, while the remaining 59\% are implemented directly by app developers. Importantly, third-party libraries are implicated in TMVs across 48.98\% of affected apps and 28.90\% of vulnerable TLS FQDNs. 

These findings underscore the significant role that third-party libraries play in propagating TLS validation vulnerabilities throughout the Android ecosystem. The widespread reuse of insecure library code not only amplifies the reach of individual vulnerabilities but also complicates remediation efforts, as a single flawed implementation can impact a large number of apps and domains. This highlights the critical need for both library maintainers and app developers to prioritize secure coding practices and to rigorously vet third-party dependencies.

\begin{table}
    \centering
    \caption{Attribution of TMVs to Responsible Parties}
    \label{tab:party_attribution_summary}
    \begin{tabular}{lccc}
        \toprule
        \textbf{Responsible Party} & \textbf{Snippets (\%)} & \textbf{Apps (\%)} & \textbf{FQDNs (\%)} \\
        \midrule
        App Developer      & 59.00 & 71.65 & 80.82 \\
        Third-Party Library & 41.00 & 48.98 & 28.90 \\
        \bottomrule
    \end{tabular}
\end{table}

\subsubject{Top Third-Party Libraries Responsible for TMVs.}
To further illustrate the impact of third-party code, Table~\ref{tab:top_3rdparty_libs} lists the top 10 third-party libraries ranked by the number of affected apps. For each library, we report the number of unique vulnerable code snippets, the number of apps in which they appear, and the number of distinct TLS FQDNs impacted. 

\begin{table}
    \centering
    \caption{Top 10 Third-Party Libraries Responsible for TMVs}
    \label{tab:top_3rdparty_libs}
    \begin{tabular}{l l c c c}
        \toprule
        \textbf{Provider} & \textbf{Library Name} & \textbf{Snippets} & \textbf{Apps} & \textbf{FQDNs} \\
        \midrule
        Aurora & JPush & 580 & 558 & 11 \\
        UMeng & UMeng+ & 477 & 414 & 9 \\
        Baidu & Map SDK & 440 & 275 & 95 \\
        Tencent & Bugly & 337 & 335 & 273 \\
        Aurora & JVerification & 288 & 156 & 1 \\
        jeasonlzy & OkGo & 84 & 62 & 104 \\
        yonyou & Native Plugins & 47 & 25 & 54 \\
        Tencent & IMSDK & 46 & 26 & 2 \\
        bokecc & Live SDK & 41 & 13 & 20 \\
        Zhichi Technologies & Sobot SDK & 29 & 21 & 10 \\
        \bottomrule
    \end{tabular}
\end{table}

Our analysis reveals that a small number of widely used libraries account for a disproportionate share of TMVs, with the top 5 libraries collectively responsible for over 81.65\% of all third-party-related vulnerabilities. This highlights the critical role of library maintainers in the security of the Android ecosystem.

\subject{Why Do Simple or Naive TLS Validation Errors Persist?}
To understand the rationale behind the prevalence of simple or naive TLS validation errors—especially in top third-party libraries—we manually reviewed the source code and documentation of the most impactful libraries. Our findings indicate both recurring themes but also previously unknown reasons. For instance, we observed a high incidence of common error patterns, such as incorrect hostname-based validation (H2-A) and checks limited to only the certificate's validity period (T2-A), within these libraries.

\subject{Summary.}
In summary, our responsible party analysis demonstrates that third-party libraries are a major source of TMVs in Android apps, both in terms of code reuse and ecosystem-wide impact. Addressing these vulnerabilities requires coordinated efforts from both app developers and library maintainers, with a particular focus on improving the security posture of widely adopted libraries. These findings highlight the critical need for improved security vetting of third-party dependencies to mitigate widespread risks.

Given TMVs, the underlying root causes and origins well understood, we then move to conduct responsible disclosure, which will be presented next in Section~\ref{sec:disclosure}.

\section{Scalable Responsible Disclosure}
\label{sec:disclosure}
Coordinating responsible disclosure for thousands of vulnerable Android applications presents significant challenges. Effective notification, communication management, and remediation tracking require a systematic and scalable approach. In this section, we detail our responsible disclosure process for TLS MitM vulnerabilities (TMVs) identified in Android apps, targeting both app developers and third-party library providers.

\subject{Contact Information Collection} To maximize disclosure coverage, we collected contact information (i.e., email address) for responsible parties using a multi-pronged strategy. First, we extracted developer email addresses from the official Google Play app store metadata. Next, we parsed app and library documentation, especially Terms of Service, using the DeepSeek V3 LLM to extract additional email addresses. 

\subject{Disclosure Protocol.} We adopt a standardized disclosure protocol. For each identified contact, we send up to three disclosure emails, spaced at least one week apart. Each email includes the app or library name, version, SHA-256 hash, MitM vulnerability types, and affected domain names. All correspondence and responses are tracked in a secure database. The disclosure email template can be found in Appendix~\ref{appendix:disclosure_template}.

\subject{Disclosure Results.}
Our responsible disclosure process is currently ongoing. By this writing, we have initiated the first phase by sending notification emails to the developers of 726 vulnerable applications. As of the submission date of this paper, no responses have been received. We will continue to execute the subsequent stages of our disclosure plan, including follow-up communications, and report the final outcomes upon completion of the process.

\subject{Lessons Learned.} Our experience highlights several key challenges in large-scale vulnerability disclosure. First, there is a pressing need for more reliable and accessible developer contact mechanisms within app stores and software repositories. Second, automated tools for extracting, verifying, and managing contact information are essential to streamline the disclosure process. Finally, stronger support from platform providers and the broader community is crucial to incentivize and facilitate timely remediation of vulnerabilities. We believe these lessons are not unique to the Android ecosystem and are equally relevant to other platforms, such as iOS.

\section{Discussion}
\label{sec:discussion}
\subject{Data and Code Release.} 
We are committed to promoting reproducibility and facilitating further research by releasing both our data and source code. Specifically, we will provide: (1) the complete source code for both \tester and \analyzer; (2) a comprehensive list of vulnerable applications identified in our study, along with their corresponding TLS MitM Vulnerabilities (TMVs); and (3) annotated vulnerable code snippets, categorized by relevant attributes. Additional supporting materials will be made available upon request.

\subject{Limitations of \tester and \analyzer.} 
While our tools, \tester and \analyzer, have proven effective in identifying TLS validation issues, several limitations remain. For \tester, the primary challenges include: (1) limited coverage of TLS flows, as the automated GUI agent may not exercise all functionalities within each application; and (2) a focus on TLS validation errors that are exploitable in practical Man-in-the-Middle (MitM) attack scenarios, potentially overlooking other types of TLS misconfigurations.

For \analyzer, the limitations include: (1) the possibility that applications may detect the presence of instrumentation and alter their behavior, thereby reducing the observable coverage of vulnerable code snippets; and (2) the inability to analyze TLS validation logic implemented in native code, as our instrumentation targets Java code exclusively. These factors collectively constrain the comprehensiveness of our vulnerability analysis.

\bibliographystyle{plain}
\bibliography{ref}
\appendices
\section{Prompt Templates for the GUI Agent in \tester}
\label{appendix:prompts}

This section details the prompt templates employed for both the general-purpose and specialized Large Language Models (LLMs) that guide the GUI agent's interaction with Android applications.

\subsection{Template for General-Purpose LLM}
The general-purpose LLM is prompted with a single, comprehensive user message. This message consolidates all necessary context, including the task description, action space, output format requirements, operational history, and the current screen's environment and element hierarchy. Figure~\ref{fig:prompt_general_llm} illustrates the complete structure of this prompt.

\subsection{Template for Specialized LLM}
The specialized LLM (UI-TARS) utilizes a multi-turn conversational format that aligns with its training regimen. The prompt begins with a system message defining the agent's persona and instructions, followed by a sequence of messages representing the interaction history. Finally, the current GUI screenshot is provided as the last user message, prompting the model to generate the next action. This structure is illustrated in Figure~\ref{fig:prompt_specialized_llm}.

\section{Impact of LLM Parameters on Performance of the GUI Agent}
\label{appendix:llm_params_eval}

We conducted experiments to evaluate the impact of different LLMs, input modalities, and action spaces on the GUI agent's performance. The setups and corresponding results are presented in Table \ref{tab:llm_setups_wide}. The following subsections describe the observations from these experiments.

\subsubject{LLM Model.}
We compared the performance of UI-TARS 1.5, Qwen2.5-VL-32B, and Gemini 2.5 Flash, with each model utilizing both UI Hierarchy and Screenshots as input and operating with a full action space. Among the three models, Gemini 2.5 Flash recorded the highest scores across all evaluated metrics. It achieved a $C_{UI}$ of 0.4515 and a $C_{TLS}$ of 0.5894. In comparison, UI-TARS 1.5 scored 0.3651 ($C_{UI}$) and 0.5405 ($C_{TLS}$), while Qwen2.5-VL-32B scored 0.3704 ($C_{UI}$) and 0.5266 ($C_{TLS}$). For novel metrics, Gemini 2.5 Flash also registered the highest performance, with a $C_{UI}^{novel}$ of 0.2945 and a $C_{TLS}^{novel}$ of 0.4449.

\subsubject{Input Modality.}
An ablation study was performed on the Qwen2.5-VL-32B model to assess the effect of input modality. We compared the performance of the agent when provided with both UI hierarchy and screenshots against using only the UI hierarchy. Removing the visual context from screenshots led to a noticeable degradation in performance. Specifically, the $C_{UI}$ score dropped from 0.3704 to 0.3351, and the $C_{TLS}$ score fell from 0.5266 to 0.5241. A similar trend was observed for novel metrics, where $C_{UI}^{novel}$ decreased from 0.1499 to 0.1182, and $C_{TLS}^{novel}$ decreased from 0.3069 to 0.2214.

\subsubject{Action Space.}
To evaluate the impact of the action space, we constrained the Qwen2.5-VL-32B model from a \texttt{Full} action space to a simplified one consisting of only \texttt{Click} and \texttt{Back} actions. This change led to a decrease in most metrics. The $C_{UI}$ score fell from 0.3704 to 0.3545, the $C_{UI}^{novel}$ score dropped from 0.1499 to 0.1182, and the $C_{TLS}^{novel}$ score decreased from 0.3069 to 0.2767. Conversely, the $C_{TLS}$ score showed a slight increase from 0.5266 to 0.5415.

\subsubject{Overall.}
In summary, our experiments demonstrate that the choice of LLM, the inclusion of multimodal inputs, and the scope of the action space are all critical factors for building a GUI agent. Among the models evaluated, Gemini 2.5 Flash demonstrated the highest performance across our established metrics. The ablation studies suggest that a multimodal input approach, which combines the UI hierarchy with screenshots, yields improved outcomes compared to relying solely on the UI hierarchy. Similarly, a comprehensive action space appears to be beneficial, affording the agent greater operational flexibility and leading to more effective task completion than a restricted action set.

\section{Prompt Template for the Vulnerable Code Classifier}
\label{appendix:classifier_prompts}

This section details the prompt templates used by the LLM in the Vulnerable Code Classifier component. We developed two main versions of the prompt, referred to as P1 and P2 in our experiments. The primary difference is that the P2 template includes an "Unknown" or umbrella category (e.g., \texttt{TU}) to handle unclassifiable cases, while the P1 template omits it.

Figure~\ref{fig:prompt_classifier} presents the general structure of the prompt template, using the classification of insecure \texttt{TrustManager} implementations as a representative example. The specific variation for the P2 template is noted within the category definitions.

\section{Responsible Disclosure Email Template}
\label{appendix:disclosure_template}

To ensure consistency and clarity across our large-scale disclosure efforts, we developed a standardized email template. This template provides developers and library maintainers with clear, actionable information regarding the identified TLS MitM vulnerabilities, including details about the affected component and specific examples of the insecure behavior. The full template used for our initial contact is presented in Figure~\ref{fig:disclosure_template}.

\begin{figure*}[p]
\begin{tcolorbox}[
  colback=white,
  colframe=black,
  boxrule=0.4pt,
  title=\textbf{Prompt Template for General-Purpose LLM},
  halign=left,
  valign=top,
  sharp corners,
  fonttitle=\bfseries
]

\textbf{Task description} \\
You are an expert familiar with automated testing of mobile applications, trying to find potential vulnerabilities through the network traffic generated by the application. Your goal is to go through all possible actions and try to trigger as diverse a range of network traffic as possible, covering as many pages as possible and making sure that critical traffic (e.g. logins, searches, comments, etc.) is triggered.

\vspace{4mm}

\textbf{Action Space}
\begin{itemize}
    \item \texttt{click [x] [y]}: Click a location on the screen.
    \item \texttt{long\_click [x] [y]}: Long-click a location on the screen.
    \item \texttt{type [text]}: Type text in the currently focused input box.
    \item \texttt{scroll [x] [y] [direction]}: Slide at a position in a given direction ('up', 'down', 'left', 'right').
    \item \texttt{drag [start\_x] [start\_y] [end\_x] [end\_y]}: Drag from a start point to an end point.
    \item \texttt{back}: Press the return key.
    \item \texttt{wait}: Wait for a period of time, often for a page to load.
    \item \texttt{finish}: End the test early if the task cannot be completed.
\end{itemize}

\vspace{4mm}

\textbf{Output Format}
\begin{verbatim}
Thoughts: <Thinking about the current decision>
Action: <Action> [parameter 1] [parameter 2] ...
\end{verbatim}

\vspace{4mm}

\textbf{Sample Output}
\begin{verbatim}
Thoughts: XXX
Action: click 100 80
\end{verbatim}

\vspace{4mm}

\textbf{Task Requirements}
\begin{itemize}
    \item You must strictly follow the output format and use plain text output (no markdown).
    \item You need to use a short paragraph in \texttt{Thoughts} to briefly describe your decision-making process (under 100 words, no line breaks).
    \item The \texttt{Action} line must strictly follow the action space and its parameter list; separate parameters with a space.
    \item Absolutely prohibit any output other than the required \texttt{Thoughts} and \texttt{Action}.
    \item The selected coordinates must be within the screen bounds.
    \item You can calculate coordinates from the element's \texttt{bounds} attribute, which is in the format \texttt{[left, top][right, bottom]}.
\end{itemize}

\textbf{History Operations}\\
\textit{[History of previous actions and thoughts]}

\vspace{4mm}

\textbf{Environment Information} \\
\begin{itemize}
    \item Screen size: \textit{[Screen Width]} x \textit{[Screen Height]}
    \item Current Activity: \textit{[Current Activity Name]}
\end{itemize}

\vspace{4mm}

\textbf{Element Information of Current Activity}\\
\textit{[Hierarchy of UI elements on the current screen]}
\end{tcolorbox}
\caption{Structure of the prompt template used for the general-purpose LLM-based GUI agent.}
\label{fig:prompt_general_llm}
\end{figure*}

\begin{figure*}[p]
\begin{tcolorbox}[
  colback=white,
  colframe=black,
  boxrule=0.4pt,
  title=\textbf{Prompt Template for Specialized LLM},
  halign=left,
  valign=top,
  sharp corners,
  fonttitle=\bfseries
]

\textit{The prompt is structured as a conversation with alternating roles.}

\vspace{4mm}

\textbf{Role: User (Initial System Instructions)}

You are a GUI agent. You are given a task and your action history, with screenshots. You need to perform the next action to complete the task.

\vspace{2mm}

\textbf{Output Format}
\begin{verbatim}
Thought: ...
Action: ...
\end{verbatim}

\vspace{2mm}

\textbf{Action Space}
\begin{itemize}
    \item \texttt{click(point='<point>x1 y1</point>')}
    \item \texttt{long\_press(point='<point>x1 y1</point>')}
    \item \texttt{type(content='...')}
    \item \texttt{scroll(point='<point>x1 y1</point>', direction='...')'}
    \item \texttt{drag(start\_point='<point>x1 y1</point>', end\_point='<point>x2 y2</point>')}
    \item \texttt{press\_back()}
    \item \texttt{wait()}
    \item \texttt{finished(content='...')}
\end{itemize}

\vspace{2mm}

\textbf{User Instruction} \\
You are an expert familiar with automated testing of mobile applications, trying to find potential vulnerabilities through the network traffic generated by the application. Your goal is to go through all possible actions and try to trigger as diverse a range of network traffic as possible, covering as many pages as possible and making sure that critical traffic (e.g. logins, searches, comments, etc.) is triggered.

\vspace{2mm}
\noindent\rule{\linewidth}{0.4pt}
\vspace{2mm}

\textbf{Role: Assistant (Interaction History 1)}
\vspace{2mm}

\texttt{Thought: \textit{[Example thought from a previous step]}}\\
\texttt{Action: \textit{[Example action from a previous step]}}

\vspace{2mm}
\noindent\rule{\linewidth}{0.4pt}
\vspace{2mm}

\textbf{Role: Assistant (Interaction History 2)}
\vspace{2mm}

\texttt{Thought: \textit{[Another example thought from a previous step]}}\\
\texttt{Action: \textit{[Another example action from a previous step]}}

\vspace{2mm}
\noindent\rule{\linewidth}{0.4pt}
\vspace{2mm}

\textbf{Role: User (Current Observation)}
\vspace{2mm}

\textit{[An image of the current GUI screenshot is provided here.]}

\end{tcolorbox}
\caption{Conversational structure of the prompt template used for the specialized LLM-based GUI agent.}
\label{fig:prompt_specialized_llm}
\end{figure*}

\begin{table*}[t]
    \centering
    \caption{Performance of GUI Agents with Different LLM Setups.}
    \label{tab:llm_setups_wide}
    \begin{threeparttable}
    \begin{tabular}{lllcccc}
    \toprule
    \textbf{LLM Model} & \textbf{Input Modality} & \textbf{Action Space} & $C_{UI}$ & $C_{UI}^{novel}$ & $C_{TLS}$ & $C_{TLS}^{novel}$ \\
    \midrule
    Qwen2.5-VL-32B & UI Hierarchy & Full & 0.3351 & 0.1182 & 0.5241 & 0.2214 \\  
    Qwen2.5-VL-32B & UI Hierarchy, Screenshots & Click, Back & 0.3545 & 0.1182 & 0.5415 & 0.2767 \\  
    Qwen2.5-VL-32B & UI Hierarchy, Screenshots & Full & 0.3704 & 0.1499 & 0.5266 & 0.3069 \\  
    UI-TARS 1.5 & UI Hierarchy, Screenshots & Full & 0.3651 & 0.1852 & 0.5405 & 0.2762 \\  
    Gemini 2.5 Flash & UI Hierarchy, Screenshots & Full & \textbf{0.4515} & \textbf{0.2945} & \textbf{0.5894} & \textbf{0.4449} \\  
    \bottomrule
    \end{tabular}
    \begin{tablenotes}
    \item [a] For all setups, the interaction wait time (WT) is 4 seconds.
    \end{tablenotes}
    \end{threeparttable}
\end{table*}

\begin{figure*}[p]
    \centering
    \begin{tcolorbox}[
      colback=white,
      colframe=black,
      boxrule=0.4pt,
      title=\textbf{Template for Responsible Disclosure Email},
      halign=left,
      valign=top,
      sharp corners
    ]
    \textbf{Subject: Security Risk Disclosure: Man-in-the-Middle Attack Vulnerabilities in \textit{[App/Library Name]}}

    \vspace{1em}
    Dear \textit{[Developer Team / Maintainer]},

    \vspace{1em}
    We are a security research team from \textit{[Institution]} led by Professor \textit{[Lead Researcher Name]}. During our security analysis of the Android component, \textbf{\textit{[App/Library Name]}}, we identified vulnerabilities in its TLS certificate validation process. These issues could expose the application to man-in-the-middle (MITM) attacks, potentially compromising user data.

    To assist your team in addressing these issues, we have summarized our findings below. Please let us know if you require further technical details.

    \vspace{1em}
    \noindent\textbf{Vulnerability Summary}\\[0.5em]
    \noindent Affected Component Information:\\
    \quad Version: \texttt{\textit{[Version Number]}}\\
    \quad SHA-256: \texttt{\textit{[SHA-256 Checksum]}}\\

    \vspace{1em}
    Identified Vulnerabilities:\\
    \begin{enumerate}
        \item \textbf{Failure to Validate Certificate Authority (CA)}\\
        \textit{Example:} HTTPS communication with \texttt{\textit{[example.vulnerable.domain]}} does not properly validate the certificate's issuing CA.\\

        \item \textbf{Domain Name Mismatch}\\
        \textit{Example:} The component accepts a valid certificate for an incorrect domain when communicating with \texttt{\textit{[another.vulnerable.domain]}}.
    \end{enumerate}

    We look forward to your response and are available to assist in the remediation process.

    \vspace{1em}
    Best regards,

    \vspace{1em}
    \textit{[Research Team Name]}
    \end{tcolorbox}
    \caption{The email template used for responsible disclosure of TLS MitM vulnerabilities.}
    \label{fig:disclosure_template}
\end{figure*}

\begin{figure*}[p]
\begin{tcolorbox}[
  colback=white,
  colframe=black,
  boxrule=0.4pt,
  title=\textbf{Prompt Template for the Vulnerable Code Classifier},
  halign=left,
  valign=top,
  sharp corners,
  fonttitle=\bfseries
]

\textbf{Task} \\
You are a professional information security researcher, and you need to thoroughly analyze the code snippet provided by the user that may be benign or contain SSL vulnerabilities.

Refer to the categories provided below to determine whether the code has vulnerabilities and the category of the vulnerabilities:

\vspace{4mm}

\textbf{Vulnerability Categories (TrustManager Example)}
\begin{itemize}
    \item \texttt{T0}: Secure TrustManager
    \item \texttt{T1}: Empty TrustManager
    \item Non-Empty but Insecure TrustManager
    \begin{itemize}
        \item \texttt{T2-A}: Only checked validity period
        \item \texttt{T2-B}: Only checked if the parameters were empty or null
        \item \texttt{T2-C}: Only checked the certificate's subject
        \item \texttt{T2-D}: Verified signature but not certificate chain
        \item \texttt{T2-E}: Ignored certificate validation exception
        \item \texttt{T2-F}: Verified certificates only under limited conditions
    \end{itemize}
    \item \textit{Note: The P2 template additionally includes the following category:}
    \begin{itemize}
        \item \texttt{TU}: Unknown, unable to determine, or the code cannot be classified into any of the above categories
    \end{itemize}
\end{itemize}

\vspace{4mm}

\textbf{Example} \\
\textit{Input}
\begin{lstlisting}[language=Java, basicstyle=\ttfamily\small, columns=flexible]
public void checkServerTrusted(X509Certificate[] p1, String p2) {
    return;
}
\end{lstlisting}
\textit{Output} \\
\texttt{T1}

\vspace{2mm}
\textit{[Additional Examples ...]}

\vspace{4mm}

\textbf{Requirements}
\begin{itemize}
    \item If the code belongs to multiple vulnerability categories at the same level, you need to output all categories separated by commas.
    \item \texttt{T2-A} / \texttt{T2-B} / \texttt{T2-C} / \texttt{T2-D} are mutually exclusive. At any time, the code will not contain more than three types of vulnerabilities.
    \item DO NOT use any format or include any additional content, \textbf{only output the classification category code}.
    \item You must strictly follow the above category definitions, and are prohibited from defining any new categories.
    \item Focus only on the method "\textit{[Method Name]}" in the class "\textit{[Class Name]}", ignore vulnerabilities in other methods.
\end{itemize}

\end{tcolorbox}
\caption{General structure of the prompt template used for the Vulnerable Code Classifier. This example shows the classification categories for insecure \texttt{TrustManager} implementations. The primary difference between the P1 and P2 templates is the inclusion of the \texttt{TU} (Unknown) category in P2.}
\label{fig:prompt_classifier}
\end{figure*}

\end{document}